\documentclass[journal]{IEEEtran}
\usepackage{cite}
\usepackage{amsmath,amssymb,amsfonts}
\usepackage{algorithmic}
\usepackage{graphicx}
\usepackage{textcomp}
\usepackage{multirow}
\usepackage{array}
\usepackage{makecell}
\usepackage{xcolor}
\usepackage{subcaption}
\usepackage{booktabs}
\usepackage{tabularx}
\usepackage{tabularx}
\usepackage{booktabs}
\usepackage{multirow}
\usepackage{colortbl}

\ifCLASSINFOpdf
\else
\fi
\begin{document}
%
\title{Measurement-Based Massive MIMO Channel Characterization and Performance Evaluation at FR3 (8 and 15 GHz) Under Equal Physical Aperture}
%
%
%

\author{
\IEEEauthorblockN{
Enrui Liu\IEEEauthorrefmark{1},
Pan Tang\IEEEauthorrefmark{1}, 
Haiyang Miao\IEEEauthorrefmark{1},
Qi Zhen\IEEEauthorrefmark{1},
Jianhua Zhang\IEEEauthorrefmark{1},
}
\IEEEauthorblockA{\IEEEauthorrefmark{1}Beijing University of Posts and Telecommunications, Beijing, China\\
Email: \{liuenrui, tangpan27, hymiao, zq2024018002, jhzhang\}@bupt.edu.cn}
}

%
%

\markboth{IEEE Transactions on XXX}%
{Liu \MakeLowercase{\textit{et al.}}: FR3 Massive MIMO Characterization}
%



\maketitle

\begin{abstract}
With the commercialization of sixth-generation (6G) wireless systems, Frequency Range 3 (FR3) has emerged as a key spectrum candidate. Bands such as 7125–8400 MHz and 14.8–15.3 GHz were identified as primary focus areas at the World Radiocommunication Conference 2023 (WRC-23), as these frequency bands offer unique advantages for achieving wide-area, high-capacity coverage. However, in practical deployments, system performance is restricted by the physical aperture constraints of base station antennas, and measurement-based comparisons of channel characteristics and system performance under these conditions require further study. Therefore, this paper conducts comprehensive measurements in Urban Macro (UMa) scenarios using a unified dual-band channel sounding platform. A cross-frequency comparison is implemented under the condition of equal antenna array area, and channel characteristics are compared. The results reveal the sparse characteristics of high-frequency bands in both the delay and spatial domains. Subsequently, based on the core constraint of ``equal physical array aperture,'' this paper integrates theoretical and empirical data to systematically evaluate cross-band coverage capability, spectral efficiency (SE). The research findings show that: regarding coverage, although 15 GHz can theoretically integrate four times the number of elements (128 elements) to offset propagation loss, a residual coverage deficit of approximately 3.0 dB remains in rigorous environments such as cell edges according to measurements; however, regarding capacity, 15 GHz successfully overcomes channel sparsity by the increased element count within the same aperture, with its spectral efficiency significantly outperforming the 8 GHz baseline. Furthermore, evaluating various array topologies (e.g., $1\times32$, $2\times16$, and $4\times8$) reveals that system performance is largely insensitive to specific geometrical layouts for a fixed number of elements. In summary, system performance is governed by the balance between propagation characteristics and hardware-enabled gain, establishing a key theoretical foundation for guiding the spatial-domain system design of next-generation FR3 bands and resolving engineering implementation challenges for 6G base stations.
\end{abstract}


%
\IEEEpeerreviewmaketitle

\section{Introduction}
%
%
%
%
\IEEEPARstart{W}{i}reless spectrum has emerged as a strategic resource for establishing competitive advantages in future 6G mobile communication systems. At WRC-23 \cite{wrc23}, upper mid-band and high-frequency ranges, represented by 7125–8400 MHz and 14.8–15.3 GHz, were formally identified as key candidate bands for 6G \cite{Rappaport5}. These FR3 bands offer unique strategic value for enabling wide-area high-capacity coverage. Compared to conventional sub-6 GHz bands, they provide 5–10× larger contiguous bandwidth, while, in contrast to millimeter-wave frequencies, they exhibit significantly lower propagation loss (by approximately 15–30 dB per 100 meters), enabling extended coverage ranges of several kilometers\cite{1,2}. However, as the carrier frequency increases, the underlying propagation mechanisms undergo fundamental changes. The growth in path loss cannot be characterized by simple linear-relationship, and small-scale channel properties—including multipath richness, delay and angular spreads, as well as the Rician K-factor—exhibit complex frequency-dependent behaviors. These effects introduce significant challenges to channel measurement and modeling. Therefore, comprehensive measurement and characterization of propagation channels in representative scenarios at FR3 candidate bands are essential for understanding the underlying electromagnetic evolution at higher frequencies.

However, investigating propagation mechanisms alone is insufficient to evaluate the true potential of practical 6G base stations. To translate the FR3 band's theoretical promise into real-world Massive MIMO deployments, cross-band evaluations require a realistic hardware baseline. Since actual base station panels are strictly bounded by physical size, we must adopt a fixed-aperture constraint. This constraint triggers a fundamental trade-off: scaling from 8 GHz to 15 GHz worsens propagation impairments (e.g., higher path loss and sparsity) but simultaneously shrinks the wavelength, enabling the integration of significantly more antenna elements within the same footprint. This raises a critical question: under a fixed physical aperture, can the enhanced spatial degrees of freedom from ultra-dense arrays outweigh the inherent high-frequency propagation disadvantages? Answering this is crucial for understanding the massive MIMO coverage–capacity trade-off and guiding future FR3 system design.

\subsection{Related Works and Motivation}

Building upon the above research, a large body of measurement studies on FR3 (mid-band) channel characteristics has been conducted across typical scenarios, including UMa, urban micro-cell (UMi), and indoor hotspot (InH) environments. These studies, based on wideband or multi-frequency measurement methodologies, have systematically characterized the propagation mechanisms of the FR3 band. Existing results generally indicate that, in terms of large-scale fading, the path loss exponent under LoS and NLoS conditions typically ranges from 2.0 to 3.5 \cite{Miao2,Rappaport1,Rappaport6}. Compared with Sub-6 GHz bands, this reflects stronger propagation attenuation, while still being lower than that observed in millimeter-wave bands, thereby demonstrating the transitional nature of the FR3 band. In addition, penetration loss measurements across various building materials within the 8–24 GHz range further reveal that signal penetration capability significantly degrades as frequency increases, which is mainly attributed to enhanced reflection and increased material absorption at higher frequencies \cite{liuenrui}. In terms of small-scale channel characteristics, existing studies include MIMO measurements at 6 \cite{Miao4} and 13 GHz \cite{tangpanharden} in UMa scenarios, indoor measurements in the 6 \cite{WeiQi} and 11 GHz \cite{Kim}, as well as multi-band measurements covering 6 to 24 GHz across UMa, UMi, and O2I scenarios \cite{Miao3}, have systematically analyzed key parameters such as delay spread, angular spread, and multipath structure evolution. The results consistently show that, with increasing frequency, both delay spread and angular spread exhibit an overall decreasing trend \cite{Rappaport4}, while the number of resolvable multipath components is significantly reduced, indicating a transition of the channel from a rich-scattering structure to a sparse one \cite{Miao}. Overall, the FR3 band simultaneously exhibits higher propagation loss and a more sparse multipath structure, forming a transitional regime between the rich-scattering Sub-6 GHz channels and the highly directional millimeter-wave channels \cite{Violette}. This not only alters the statistical characteristics of the channel but also poses new challenges for system design.

At the system level, a substantial amount of research has further investigated the coverage performance and spectral efficiency across different frequency bands\cite{Mizmizi}. For instance, in multi-band Massive MIMO systems, studies have shown that, through beamforming techniques, high-frequency systems can achieve received power levels comparable to or even higher than those of low-frequency systems under certain conditions\cite{Health}. Moreover, coverage studies on millimeter-wave and FR3 bands indicate that, despite the higher path loss at high frequencies, beamforming gain provided by large-scale antenna arrays can significantly improve edge coverage performance. In terms of spectral efficiency, analyses based on stochastic channel models or ideal i.i.d. assumptions suggest that system capacity can scale approximately linearly with spatial degrees of freedom as the antenna size increases\cite{BjornsonMassiveMIMO}. However, such results often rely on idealized propagation conditions, including rich scattering environments and fully coherent combining assumptions, and therefore their applicability in practical measured channels still requires further validation.

Furthermore, in practical wireless system deployments, the physical size of antenna arrays is typically constrained by site space, structural limitations, and engineering implementation conditions \cite{Peschiera}, making it infeasible to arbitrarily scale the array size with frequency. Therefore, compared with idealized array scaling analysis, the fixed physical aperture constraint becomes a more realistic design consideration. Under this constraint, the relationship among frequency, array size, and system performance becomes more complex. On one hand, based on the half-wavelength spacing principle, the number of antenna elements scales with frequency according to $N \propto f^2$, allowing high-frequency systems to accommodate more antenna elements within the same physical area, thereby providing higher array gain and channel capacity \cite{Bjornson1}. On the other hand, the free-space path loss also increases quadratically with frequency, while practical channels additionally suffer from more severe penetration loss, reduced scattering, and multipath sparsity \cite{Rappaport2}. Therefore, although array gain can theoretically compensate for the propagation loss caused by frequency increase, it remains unclear whether such “array compensation” can fully offset the “propagation degradation” in realistic environments.

In summary, although existing studies have achieved substantial progress in characterizing multi-band channel properties and system performance, several key limitations remain. First, most system-level analyses are based on inconsistent array configurations or idealized channel models, making it difficult to fairly reflect the intrinsic differences across frequency bands. More importantly, in the FR3 band—which simultaneously exhibits “propagation degradation” and “array scaling potential”—the intrinsic trade-off between coverage performance and spectral efficiency under a fixed physical aperture constraint is still not well understood. Therefore, it is necessary to conduct systematic measurement-based studies under a unified physical aperture constraint to jointly analyze channel characteristics, coverage performance, and spectral efficiency, in order to reveal the true coupling relationship among frequency, propagation mechanisms, and array scaling.

\subsection{Contributions}

This paper investigates the performance evolution of FR3 Massive-MIMO systems under practical physical aperture constraints through a measurement-driven framework. The main contributions are summarized as follows:

\begin{enumerate}

\item A dual-band FR3 massive-MIMO measurement campaign is conducted in a typical UMa scenario at 8~GHz and 15~GHz using a unified channel sounding platform. By constructing arrays with identical physical apertures, a physically consistent cross-frequency comparison framework is established, enabling fair evaluation of frequency scaling effects under practical deployment constraints.

\item The channel characteristics are systematically analyzed in both delay and angular domains. The results reveal a clear transition from a rich multipath structure at 8~GHz to a sparser channel representation at 15~GHz, characterized by reduced delay spread and concentrated azimuth angular dispersion, while the elevation spread remains largely constrained by geometry.

\item The coverage performance under equal physical aperture constraints is quantitatively evaluated. The results show that although higher frequencies incur increased propagation loss and reduced multipath support, the larger number of antenna elements within the same aperture enables stronger coherent combining, partially compensating for coverage degradation. However, a non-negligible performance gap persists in low-power regions, indicating the fundamental limitation of aperture-based compensation in challenging propagation conditions.

\item The spectral efficiency performance is further investigated using the measured channel data. The results demonstrate that, despite the reduced multipath richness at higher frequencies, the increased spatial resolution of the electrically larger array significantly improves the ability to resolve and exploit the channel structure. This leads to more efficient utilization of spatial degrees of freedom and enables the 15~GHz system to achieve superior spectral efficiency compared to the 8~GHz baseline.

\end{enumerate}

\subsection{Organization}

The remainder of this paper is organized as follows. Section II introduces the FR3 channel measurement platform and the measurement environments. Section III presents the measured channel characteristics at 8GHz and 15GHz and analyzes their propagation and spatial properties. Section IV and V investigates the coverage performance and spectral efficiency of the two systems under equal physical aperture constraints. Finally, Section VI concludes the paper.
\section{Channel Measurement Platform and Environment}
\subsection{Measurement Platform}
\begin{figure}[!ht]
\begin{subfigure}{\columnwidth}
    \centering
    \includegraphics[scale=0.5]{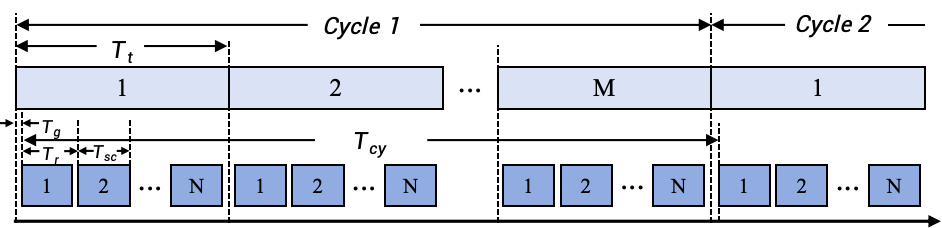}
    \caption{The principle of time-sequential switching}
    \label{fig:principle}
\end{subfigure}
\begin{subfigure}{\columnwidth}
    \centering
    \includegraphics[scale=0.25]{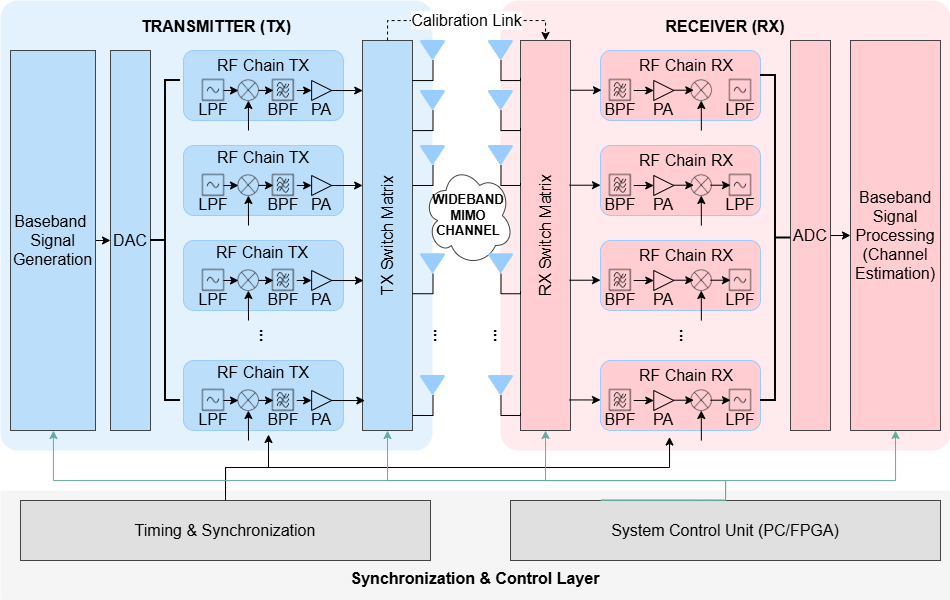}
    \caption{The measurement platform}
    \label{fig:platform}
\end{subfigure}
\caption{Schematic diagram of measurement platform.}
\label{Fig.1}
\end{figure}

Our measurement platform is designed based on the principle of time-division multiplexing (TDM), with the overall system architecture and working principle illustrated in Fig.~\ref{Fig.1}, and the detailed measurement configurations summarized in Table~\ref{Table.1}. 

For the MIMO platform, both the Tx and Rx antenna arrays are connected to their corresponding RF chains through high-speed RF switch matrices, with the switching sequences controlled by upper-computer programming. During the transmission window of each Tx antenna, all 40 Rx antenna ports are sequentially polled while ensuring that the total switching duration remains within the channel coherence time. This enables accurate acquisition of the complete $128 \times 40$ MIMO channel responses and supports reliable channel characterization even in high-mobility scenarios. For the single-input single-output (SISO) platform, a dedicated point-to-point measurement configuration is adopted. At the transmitter, the signal is amplified by a power amplifier and fed to a horn antenna, while at the receiver, an omnidirectional antenna is connected to a spectrum analyzer through a low-noise amplifier. This configuration provides a stable and well-calibrated reference measurement, which is further used for path loss and wideband channel characterization.

\begin{table}[ht]
\centering
\caption{Measurement system parameters for SISO and MIMO platforms.}
\label{Table.1}
\begin{tabular}{lcc}
\toprule
\textbf{Parameter} & \textbf{SISO Platform} & \textbf{MIMO Platform} \\
\midrule
Center frequency & 6-18 GHz & 3--16 GHz \\
Maximum bandwidth & 2 GHz & 2 GHz \\
Measurement bandwidth & 1 GHz & 400 MHz \\
Tx channels & 1 & 128 \\
Rx channels & 1 & 64 \\
Tx antenna type & Horn & UPA \\
Rx antenna type & Omnidireactional & ODA \\
Tx antenna gain & $\geq$12 dBi & $\geq$27 dBi \\
Rx antenna gain & $\geq$3 dBi & $\geq$33 dBi \\
Angular resolution & -- & $\leq 1^{\circ}$ \\
Synchronization & GPS-disciplined clock & GPS-disciplined clock \\
\bottomrule
\end{tabular}
\end{table}

In Fig. \ref{Fig.1} (a), $T_{t}$ denotes the transmission switching duration for MIMO, i.e., the time during which the signal is maintained on the Tx antenna array; $T_{r}$ denotes the reception switching duration; $T_{cy}$ is the probing cycle time, which must satisfy the condition $T_{cy} \geq M T_t$; $T_g$ is the guard interval, which is reset to zero at the beginning of the next cycle; $T_{sc}$ represents the continuous detection duration of each Rx antenna element; and $T_s$ is the pulse period, during which each Rx antenna array performs one detection. The specified probing duration $T_t$ exactly corresponds to one detection cycle, i.e., $T_t$ = N$T_{sc}$. Since switching requires additional time in practical operation, a guard interval must be introduced. Therefore, the interval between two consecutive detection signals is defined as $T_r$, with the condition $T_r \geq T_{sc}$.

Fig.~\ref{Fig.1} (b) shows the top view of the platform. The entire measurement platform’s transmitter generates PN9 sequence codewords using an signal generator, which are then transmitted via coaxial cables to a matrix switching unit. The switches are controlled by a computer, allowing the 128 transmit channels to sequentially pass through power amplifiers and excite the antenna RF ports, thereby transmitting the signals into the measurement channel. At the receiver side, the signals captured by the antennas are first amplified by low-noise amplifiers (LNAs) and then routed through the receiving-side switches into a spectrum analyzer for acquisition and processing. The timing and synchronization of the entire platform are controlled by a GPS-disciplined Rubidium clock system, ensuring nanosecond-level synchronization between all transmit and receive channels, thus providing a reliable foundation for high-precision massive MIMO channel measurements.
\subsection{Measurement Environment}
\begin{figure}[!ht]
\centering

\begin{subfigure}[b]{0.95\columnwidth}
    \centering
    \includegraphics[width=\textwidth,height=0.24\textheight]{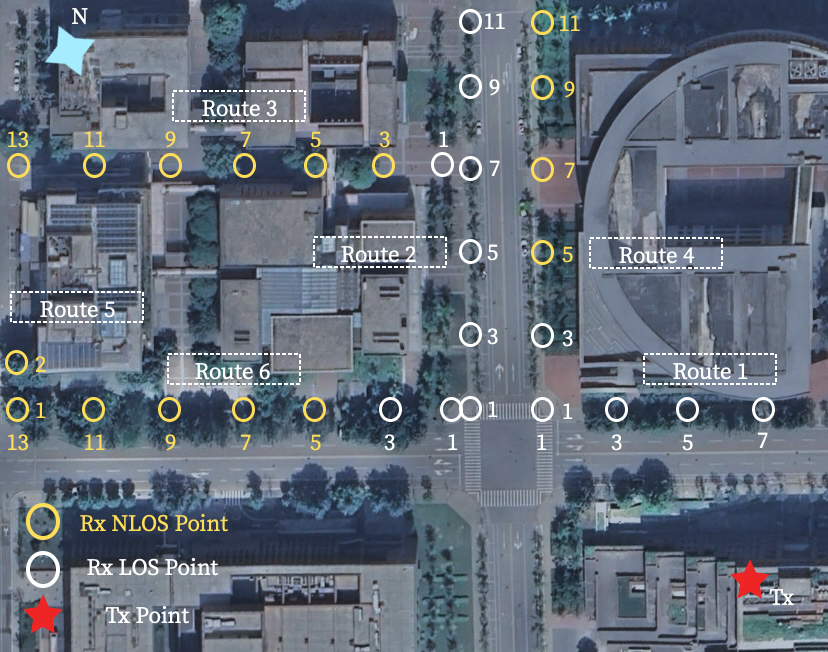}
    \caption{Measurement route and environment}
    \label{fig:route}
\end{subfigure}

\vspace{0.6em}

\begin{subfigure}[b]{0.48\columnwidth}
    \centering
    \includegraphics[width=\textwidth]{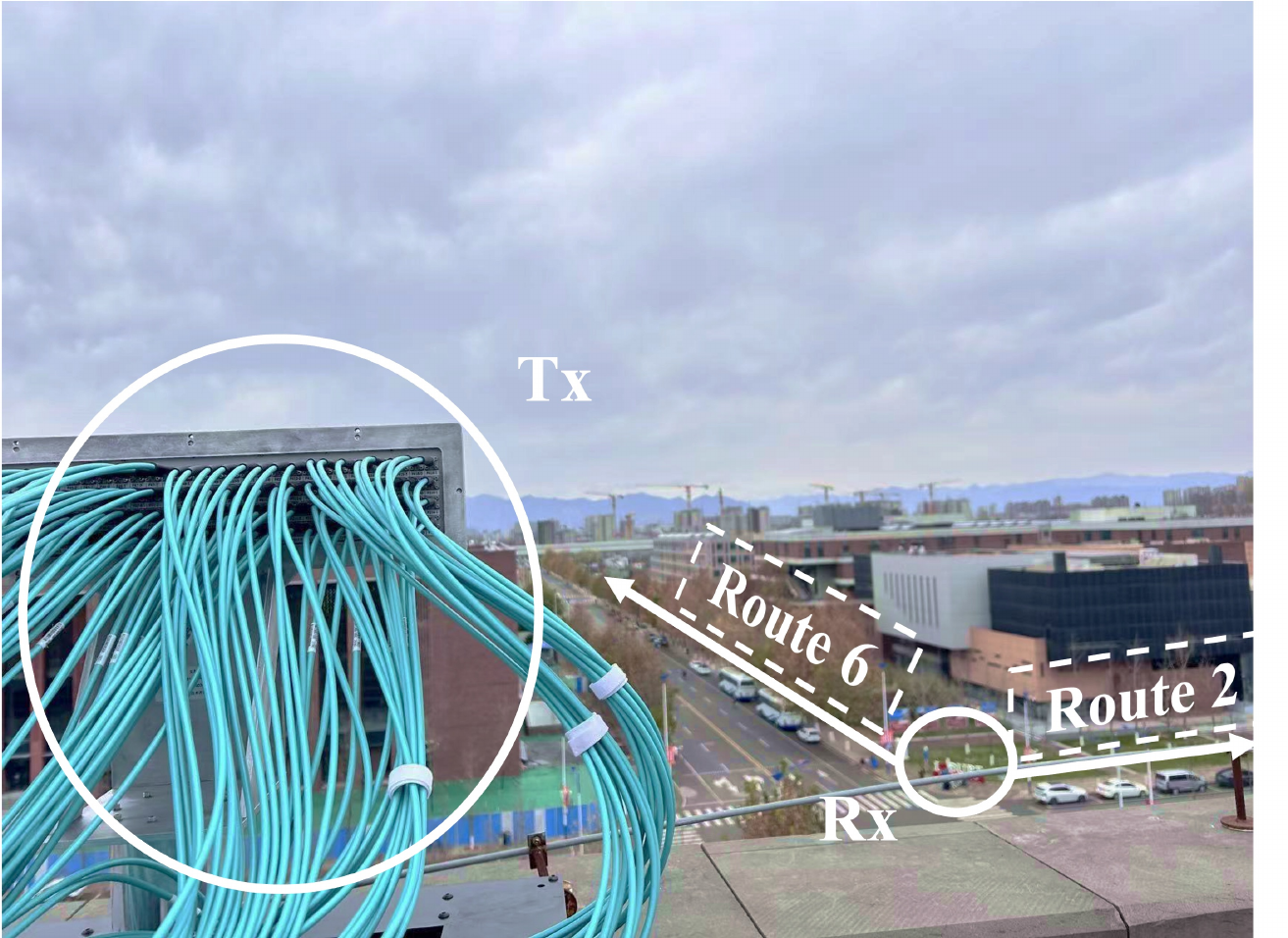}
    \caption{Transmitter}
    \label{fig:transmitter}
\end{subfigure}
\hfill
\begin{subfigure}[b]{0.48\columnwidth}
    \centering
    \includegraphics[width=\textwidth]{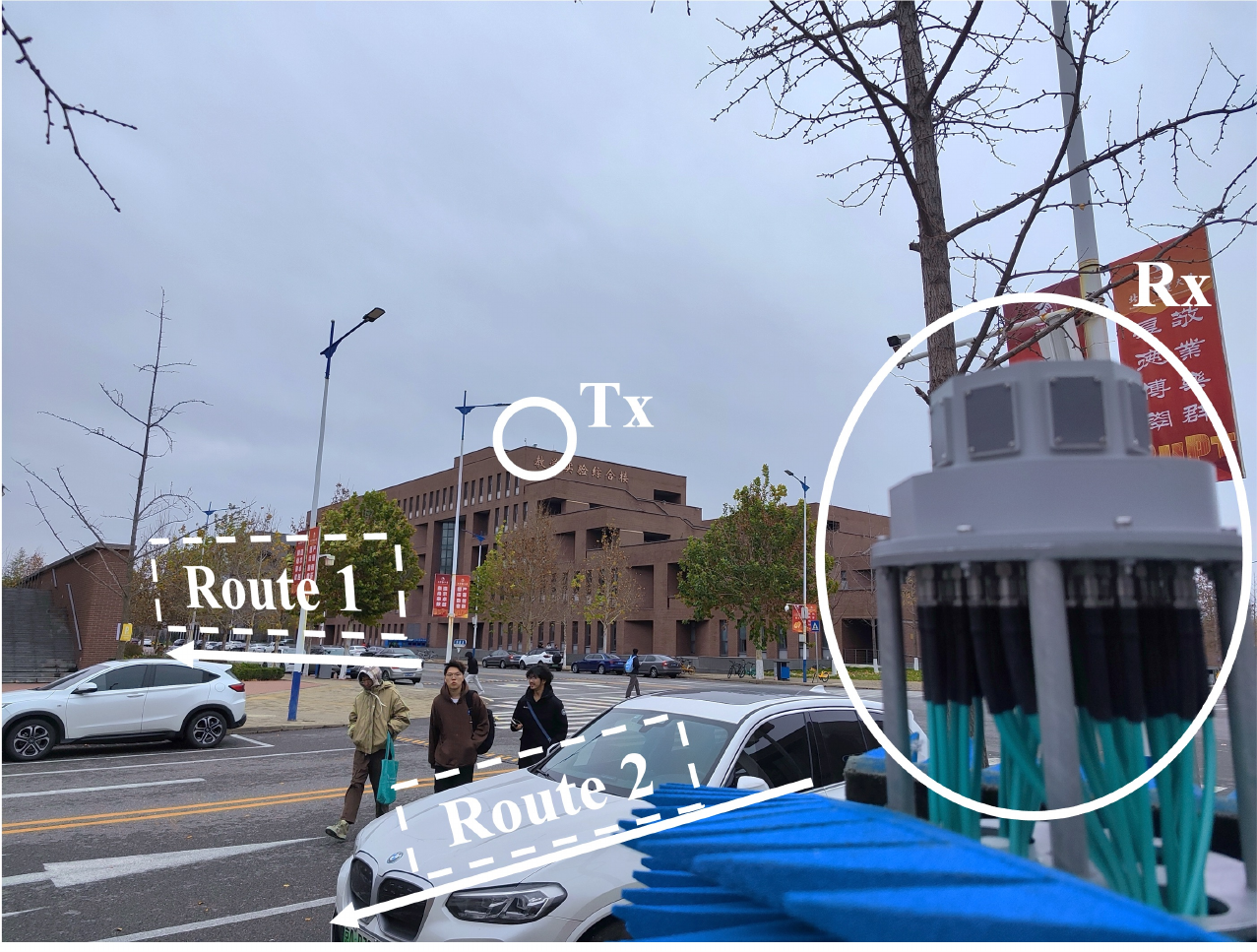}
    \caption{Receiver}
    \label{fig:receiver}
\end{subfigure}

\caption{Measurement setup in the UMa scenario, including the route layout and photographs of the transmitter and receiver setups.}
\label{fig:measurement_setup}
\end{figure}

The MIMO measurement campaign was conducted on the Shahe Campus of Beijing University of Posts and Telecommunications (BUPT). In the UMa scenario, the transmitter was deployed on the rooftop of the main building at a height of approximately 27.8~m above ground. As illustrated in Fig.~\ref{fig:measurement_setup} (a), the receiver was moved along predefined routes covering six trajectories. The corresponding propagation environment is also depicted, providing the geometric basis for subsequent channel analysis. The photos of the channel sounding system are shown in Fig.~\ref{fig:measurement_setup} (b) and Fig.~\ref{fig:measurement_setup} (c), which present the transmitter and receiver setups, respectively.

\subsection{Measurement Data Post-Processing}
Due to its excellent autocorrelation properties, simple generation process, and wide dynamic range, a PN9 sequence is employed in this work. The collected I/Q data contains the inherent response of the system. To eliminate the influence of the system on the channel characteristics and obtain the true channel impulse response (CIR), direct calibration is performed prior to measurement\cite{wang}\cite{tangpan}. The received signal can be expressed as:
\begin{equation}
y(\tau) = x(\tau)*h_{Tx}(\tau)*a_{Tx}(\tau)*h(\tau)*a_{Rx}(\tau)*h_{Rx}(\tau),
\end{equation}
where “*” denotes the convolution operation. Here, $x(\tau)$ is the transmitted PN sequence, $h_{Tx}(\tau)$ and $h_{Rx}(\tau)$ represent the responses of the transmitter and receiver RF chains, $a_{Tx}(\tau)$ and $a_{Rx}(\tau)$ denote the characteristics of the transmit and receive antennas, and $h(\tau)$ is the channel impulse response to be measured. The calibration signal is expressed as
\begin{equation}
y_{cal}(\tau) = x(\tau)*h_{Tx}(\tau)*h_{Rx}(\tau),
\end{equation}
by performing calibration, the effects of the transceiver RF chains can be removed. Specifically, the effective system response $h_{sys}(\tau) = h_{Tx}(\tau)*h_{Rx}(\tau)$ is first obtained through the direct connection measurement. Then, the received signal $y(\tau)$ is divided by the calibration response $y_{cal}(\tau)$ in the frequency domain, yielding
\begin{equation}
H(\tau) = \mathcal{F}^{-1}\left \{  {\frac{Y(f)}{Y_{cal}(f)}}\right \}  = a_{Tx}(\tau)*h(\tau)*a_{Rx}(\tau),
\end{equation}
where $\mathcal{F}^{-1}{\cdot}$ denotes the inverse Fourier transform, and $Y(f)$ and $Y_{cal}(f)$ are the Fourier transforms of $y(\tau)$ and $y_{cal}(\tau)$, respectively.

In this way, the influence of the RF front-end is eliminated, and the obtained result contains only the channel impulse response combined with the antenna characteristics. If necessary, further de-embedding of antenna effects can be applied to isolate the pure channel response $h(\tau)$.

After calibration and extraction of the channel impulse response, the high-resolution Space-Alternating Generalized Expectation-Maximization (SAGE) algorithm is employed for parameter estimation. Based on the EM framework, the SAGE algorithm alternately estimates and updates the parameters of multipath components. The optimization process of the SAGE algorithm \cite{sage} can be formulated within the EM framework. Specifically, for each subset of parameters $S$, the auxiliary function $\phi^{S}(\cdot)$ is constructed as

\begin{equation}
\phi^{S}(\theta_{S};\bar{\theta})=\int f(x|Y=y;\bar{\theta})\log_{}{f(x;\theta_{S},\bar{\theta}_{\tilde{S}})}dx-P(\theta_{S},\bar{\theta}_{\tilde{S}}),
\end{equation}
where $\theta_{S}$ represents the parameters of the current multipath component to be estimated, $\bar{\theta}$ denotes the estimate of the parameters, and $P(\cdot)$ is a penalty function introduced for regularization. The update rule of SAGE alternates between maximizing the auxiliary function with respect to $\theta_{S}$ and keeping the remaining parameters $\theta_{\tilde{S}}$ fixed, as expressed by

\begin{equation}
\theta^{i+1}_{S^i}=\arg \max_{\theta_{S^i}} \phi^{S^i}(\theta_{S^i};\theta^i),
\end{equation}
\begin{equation}
\theta^{i+1}_{\tilde{S^i}}=\theta^i_{\tilde{S^i}}.
\end{equation}
Through iterative refinement, the algorithm converges to stable estimates of multipath parameters, enabling accurate characterization of weak and closely spaced components under noisy channel conditions.

\begin{figure}[!ht]
\centering
\includegraphics[width=1\columnwidth]{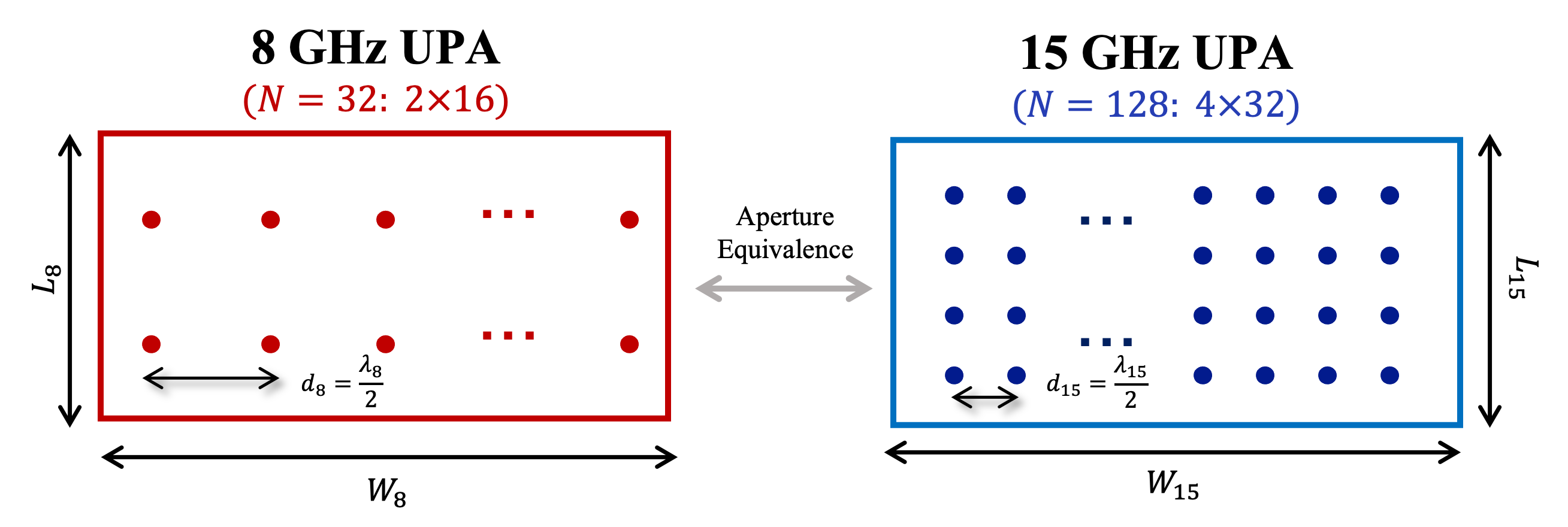}
\caption{Illustration of the fixed-aperture array configuration for 8~GHz and 15~GHz bands.}
\label{Fig.Antenna_Structure}
\end{figure}

To enable a fair comparison of channel characteristics between the 8GHz and 15GHz bands under the same total transmit power, the transmitting uniform planar array (UPA) was partitioned into subarrays to form arrays with approximately identical physical apertures. Specifically, a 32-element UPA was used at 8GHz, while a 128-element UPA was adopted at 15GHz as shown in Fig.~\ref{Fig.Antenna_Structure}.

This configuration is motivated by the half-wavelength inter-element spacing used in both bands. For a UPA with $N_x$ and $N_y$ elements along the horizontal and vertical dimensions, the physical array aperture is
\begin{equation}
A_{\text{array}} = N_x N_y \left(\frac{\lambda}{2}\right)^2,
\label{Aarray}
\end{equation}
where $\lambda = c/f$ is the wavelength. Under a fixed physical aperture, the total number of elements scales inversely with wavelength squared:
\begin{equation}
N = N_x N_y \propto \frac{A_{\text{array}}}{\lambda^2}.
\label{array}
\end{equation}
Since the wavelength at 15~GHz is roughly half that at 8~GHz, increasing the number of elements by a factor of four (from 32 to 128) yields a physical aperture comparable to the 8~GHz array. Consequently, the 15~GHz (128-element) and 8~GHz (32-element) arrays exhibit nearly identical antenna apertures, enabling an aperture-normalized comparison of propagation characteristics under the same total transmit power.

\section{Measurement Results}
In this section, channel characteristics are extracted and analyzed in both large-scale and small-scale aspects under the constraint of a fixed array aperture, and then compared across different frequency bands.

\subsection{Power Delay Profiles}

\begin{figure}[t]
    \centering
    \begin{subfigure}[b]{0.24\textwidth}
        \includegraphics[width=\textwidth]{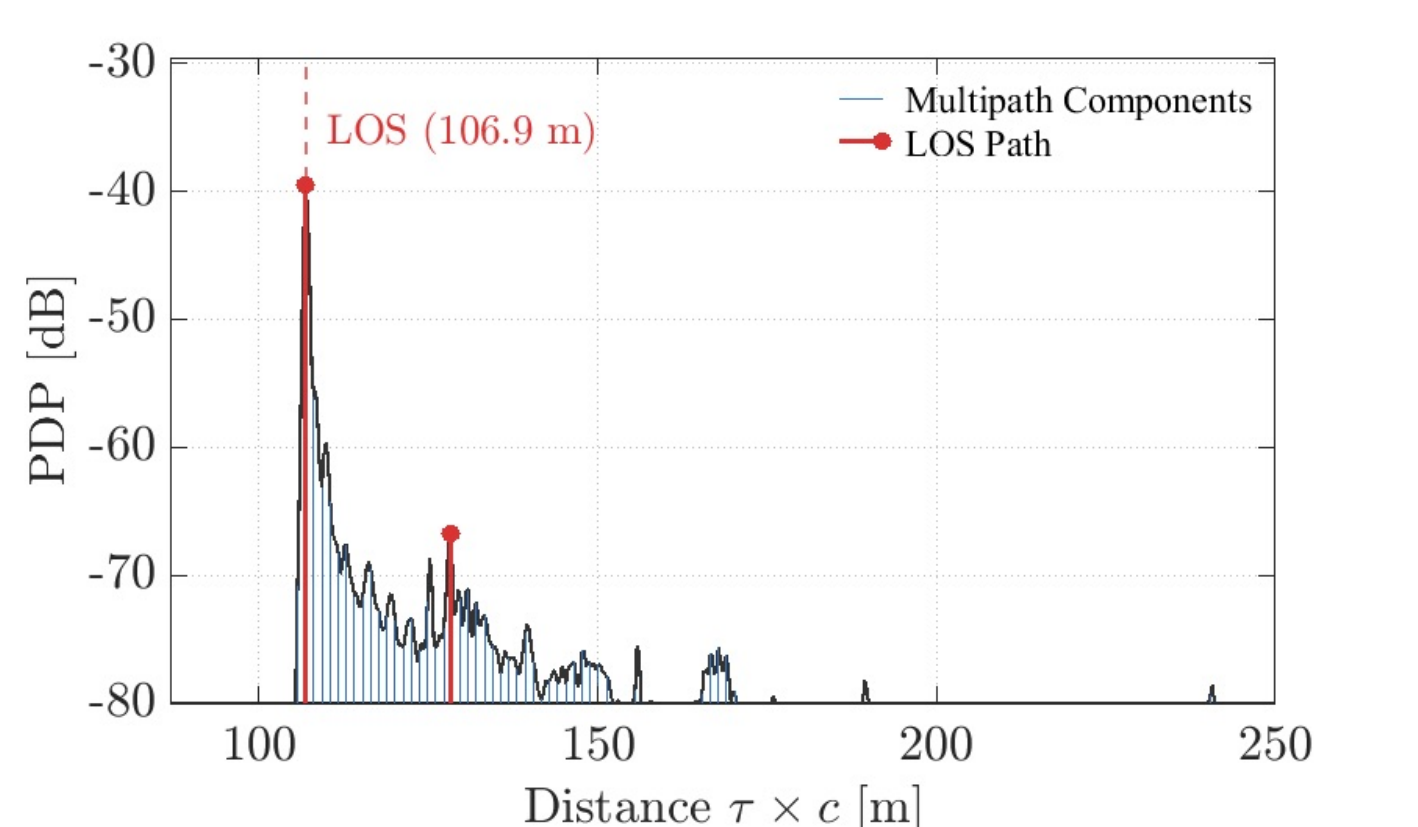}
        \caption{8 GHz, LOS}
        \label{fig:pdp_los_8}
    \end{subfigure}
    \hfill
    \begin{subfigure}[b]{0.24\textwidth}
        \includegraphics[width=\textwidth]{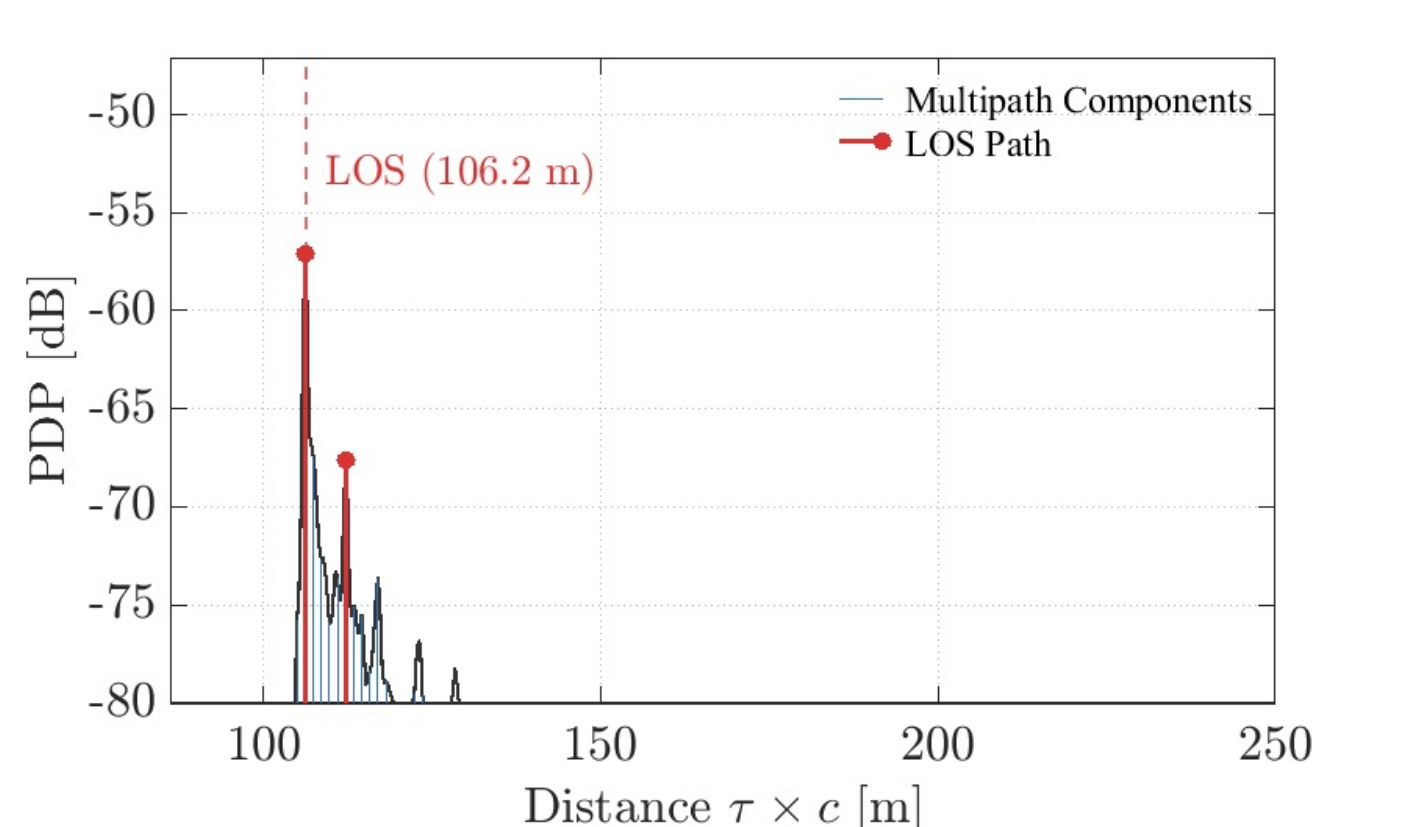}
        \caption{15 GHz, LOS}
        \label{fig:pdp_los_15}
    \end{subfigure}
    \hfill
    \begin{subfigure}[b]{0.24\textwidth}
        \includegraphics[width=\textwidth]{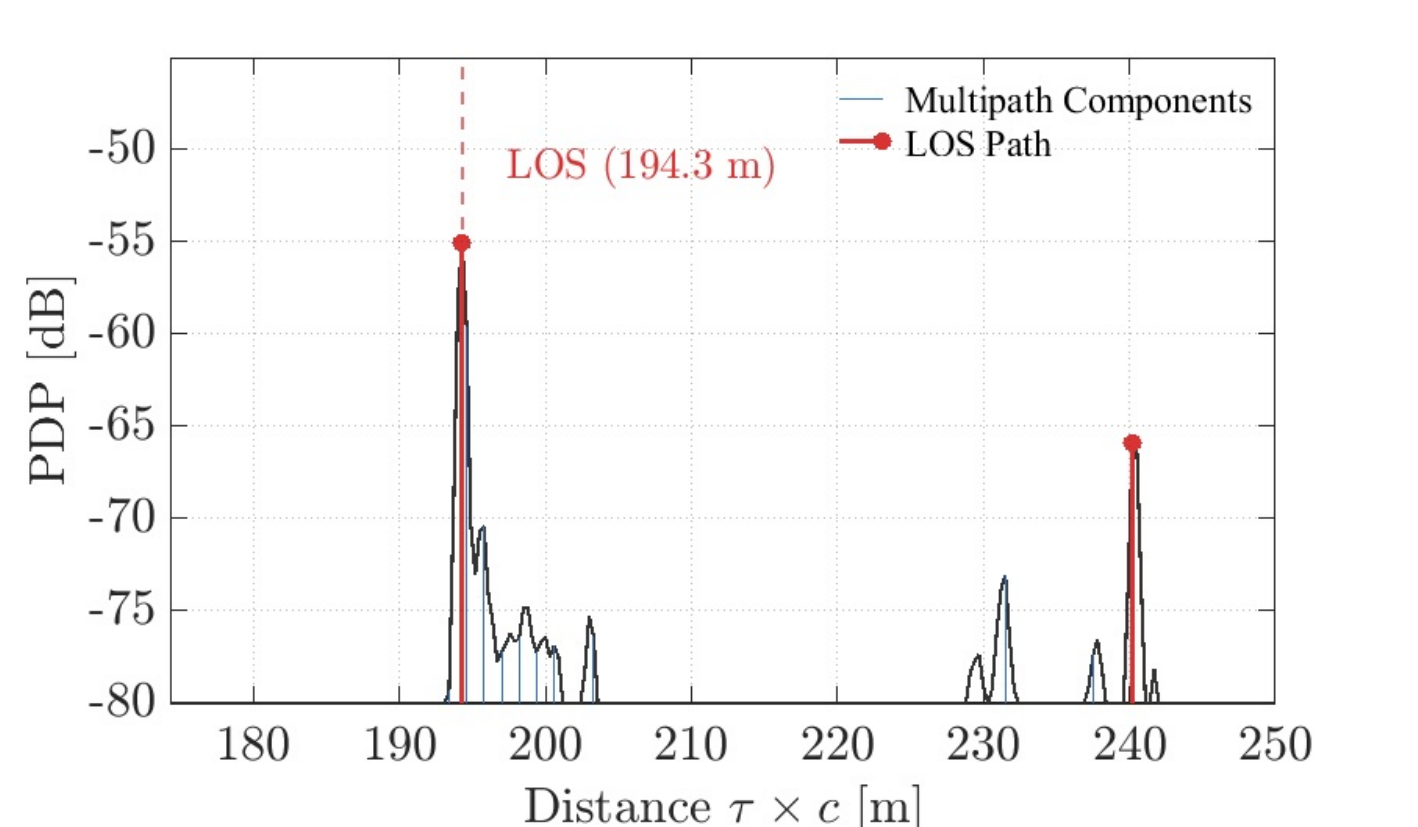}
        \caption{8 GHz, NLOS}
        \label{fig:pdp_nlos_8}
    \end{subfigure}
    \hfill
    \begin{subfigure}[b]{0.24\textwidth}
        \includegraphics[width=\textwidth]{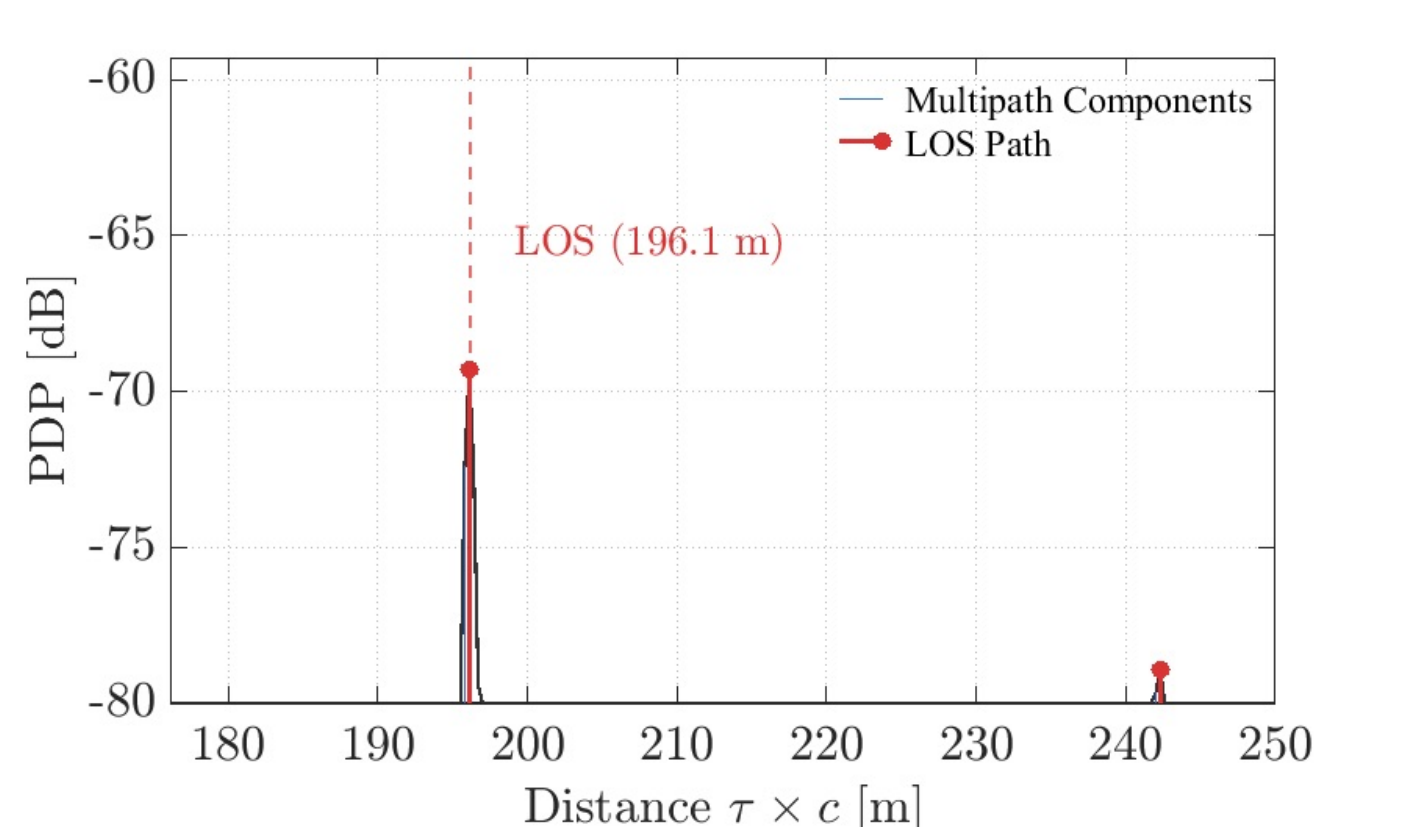}
        \caption{15 GHz, NLOS}
        \label{fig:pdp_nlos_15}
    \end{subfigure}
    \caption{Measured PDP across 8 GHz and 15 GHz bands. The red markers indicate the strongest and second-strongest multipath components.}
    \label{fig:PDP_comparison}
\end{figure}
The measured power delay profiles (PDPs) at 8~GHz and 15~GHz under LOS and NLOS conditions are shown in Fig.~\ref{fig:PDP_comparison}, where delay bins are converted to propagation distance. By leveraging the geometric layout of the environment, the dominant propagation paths can be associated with specific PDP peaks.

In the LOS scenario (route~2, point~1), the channel consists of a direct path at approximately 106~m, a ground-reflected path around 109~m, and a single-bounce reflection from a billboard at about 139~m. At 8~GHz, all these components are clearly observable, forming a rich PDP with multiple distinguishable paths and a gradual power decay across propagation orders. In contrast, at 15~GHz, while the direct path and ground reflection remain visible, the higher-delay components—particularly the billboard reflection—are significantly attenuated and may become indistinguishable from the noise floor, resulting in a simplified PDP dominated by a few strong paths. A similar trend is observed in the NLOS scenario (route~3, point~5), where the channel is mainly composed of a first-order reflection at approximately 195~m and a second-order reflection around 240~m. At 8~GHz, both components are clearly resolved, forming a structured PDP with moderate power imbalance. However, at 15~GHz, the second-order reflection is severely weakened and may fall below the noise floor, leaving only the dominant first-order path and leading to a more compact PDP with reduced delay spread. These observations indicate a clear frequency-dependent transition in the effective multipath structure. At lower frequencies (8~GHz), the channel exhibits a richer set of resolvable multipath components, where both first- and higher-order paths contribute to the received signal. As the frequency increases to 15~GHz, higher-order paths become less observable due to increased free-space attenuation and reflection losses, as well as the frequency-dependent interaction between electromagnetic waves and environmental surfaces. As a result, the channel tends to be dominated by a small number of strong components, leading to a sparser and more propagation-limited structure.

\begin{figure*}[t]
    \centering

    \begin{subfigure}[b]{0.32\textwidth}
        \includegraphics[width=\textwidth]{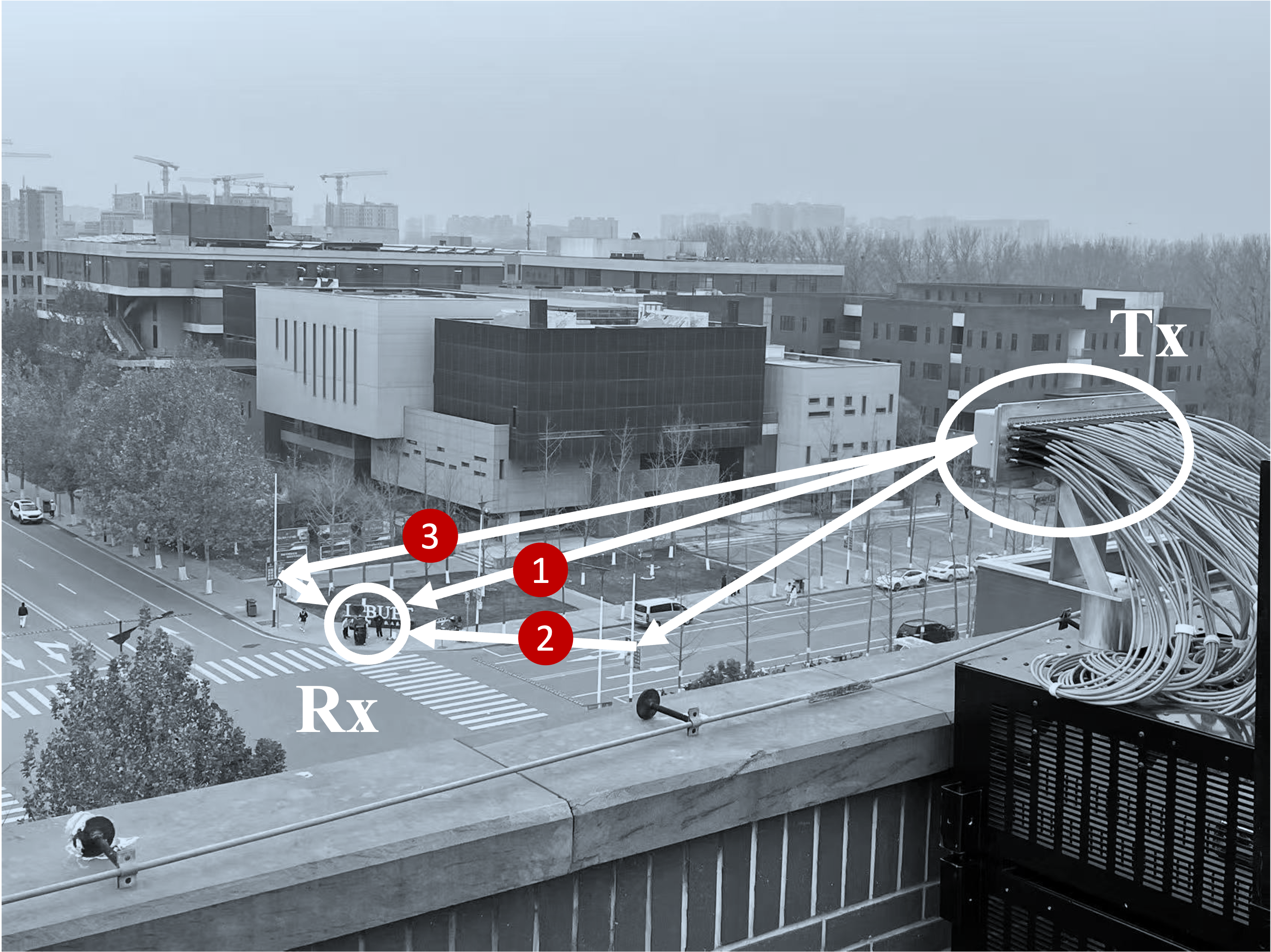}
        \caption{LOS Scenario (Environment)}
        \label{fig:los_env}
    \end{subfigure}
    \hfill
    \begin{subfigure}[b]{0.32\textwidth}
        \includegraphics[width=\textwidth]{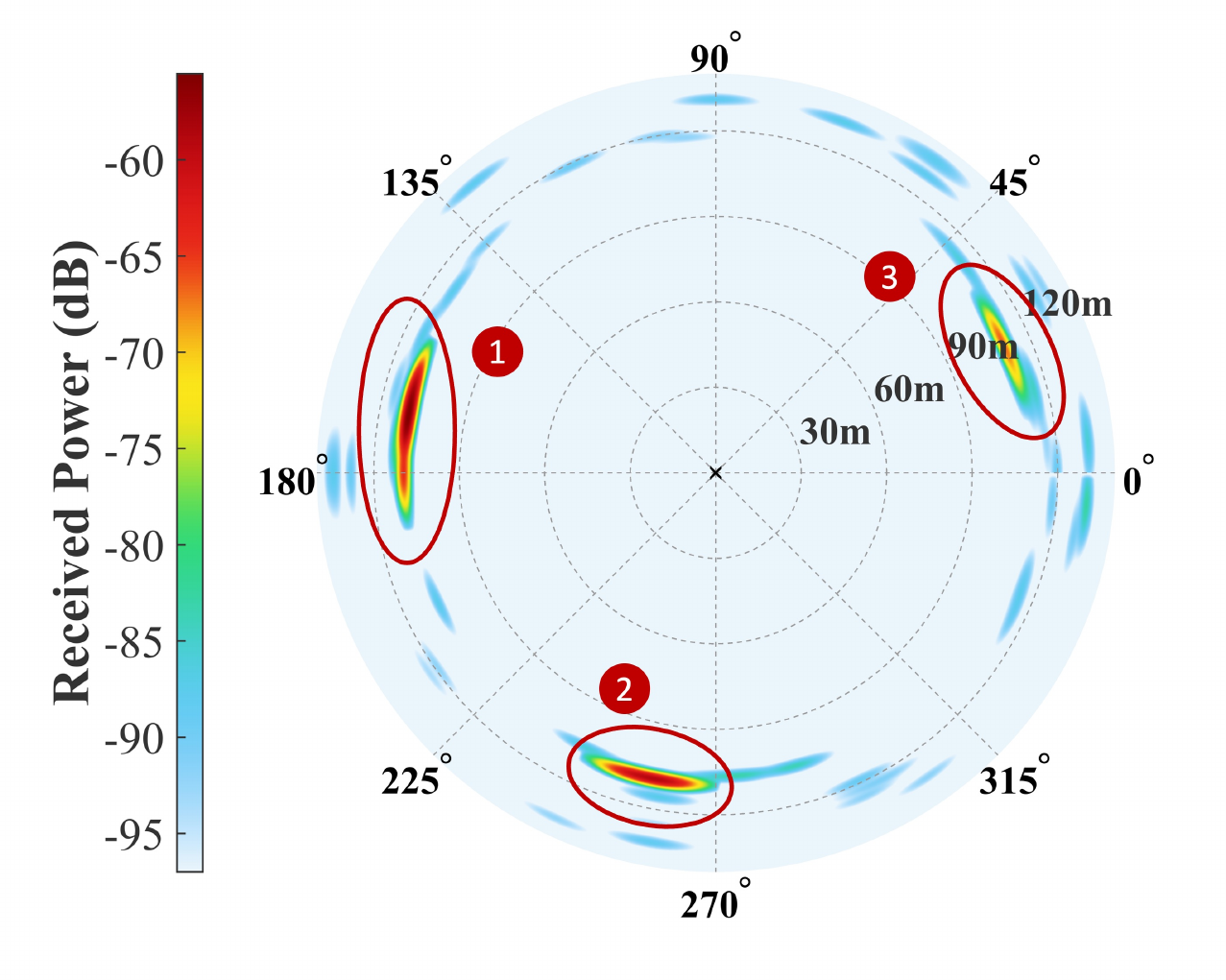}
        \caption{LOS @ 8 GHz}
        \label{fig:dpas_los_8}
    \end{subfigure}
    \hfill
    \begin{subfigure}[b]{0.32\textwidth}
        \includegraphics[width=\textwidth]{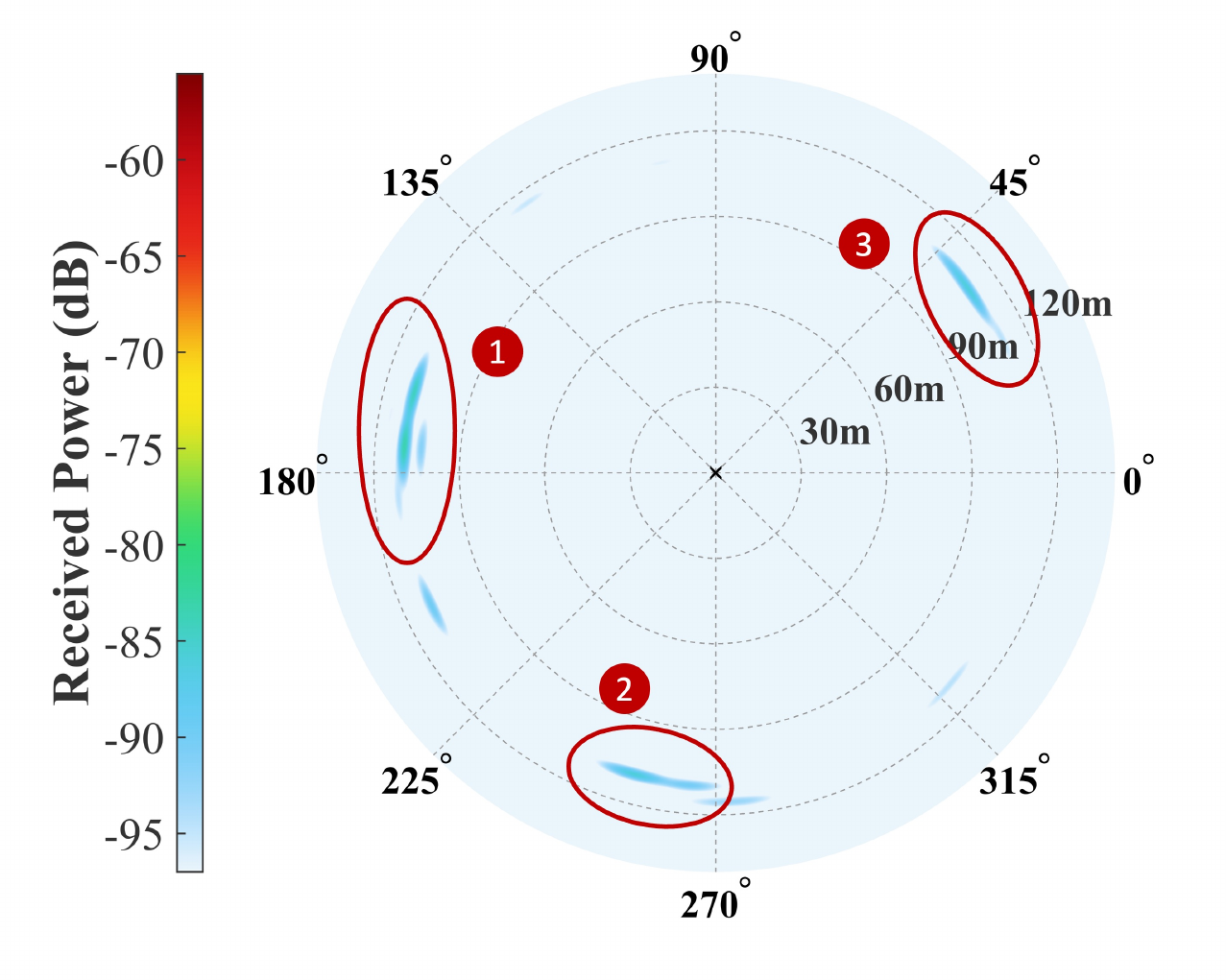}
        \caption{LOS @ 15 GHz}
        \label{fig:dpas_los_15}
    \end{subfigure}

    \vspace{0.5em}

    \begin{subfigure}[b]{0.32\textwidth}
        \includegraphics[width=\textwidth]{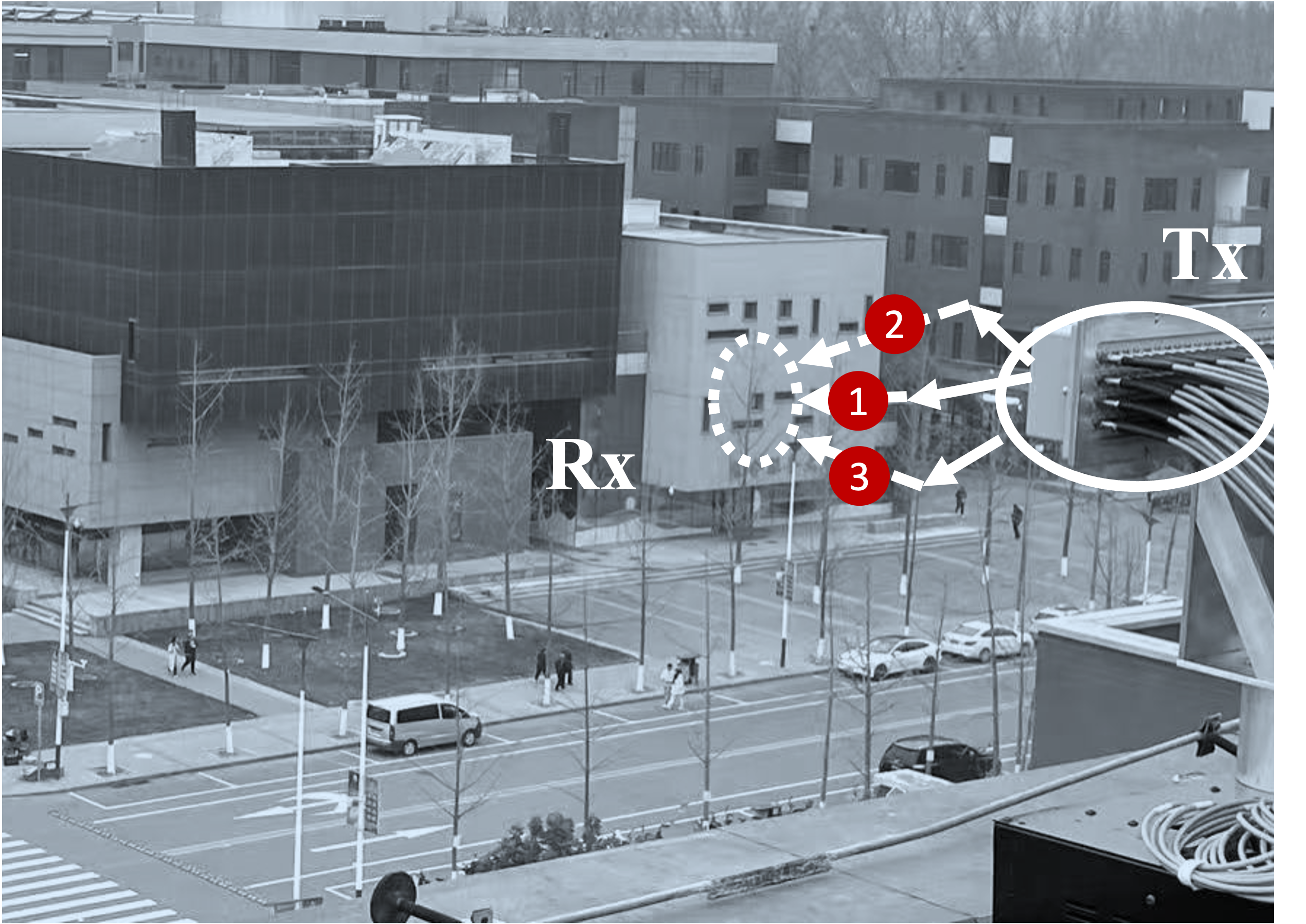}
        \caption{NLOS Scenario (Environment)}
        \label{fig:nlos_env}
    \end{subfigure}
    \hfill
    \begin{subfigure}[b]{0.32\textwidth}
        \includegraphics[width=\textwidth]{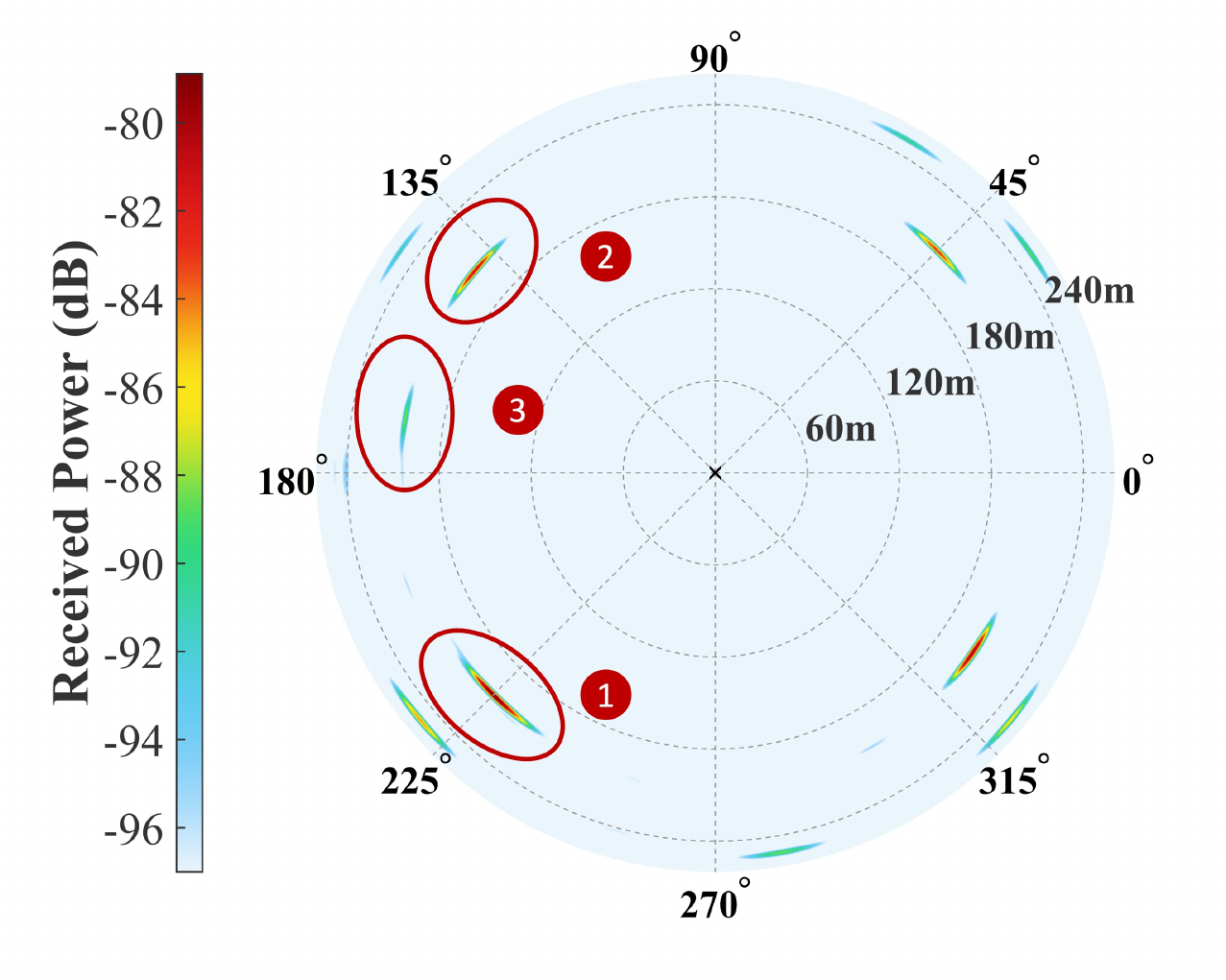}
        \caption{NLOS @ 8 GHz}
        \label{fig:dpas_nlos_8}
    \end{subfigure}
    \hfill
    \begin{subfigure}[b]{0.32\textwidth}
        \includegraphics[width=\textwidth]{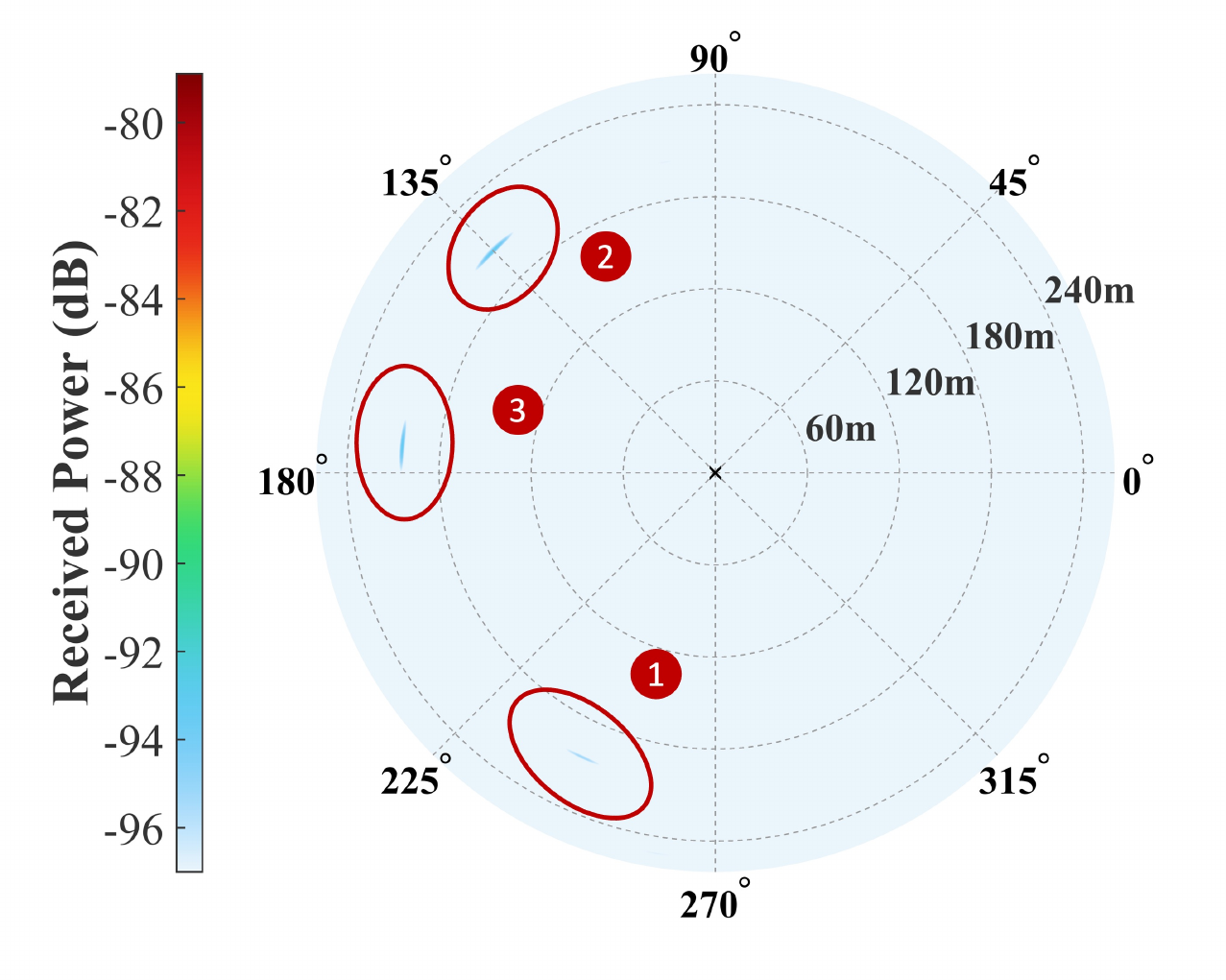}
        \caption{NLOS @ 15 GHz}
        \label{fig:dpas_nlos_15}
    \end{subfigure}

    \caption{DAPSs under equal physical aperture. The first column shows the measurement environments for LOS and NLOS scenarios, while the second and third columns present the corresponding DAPS at 8~GHz and 15~GHz, respectively.}
    
    \label{fig:DPAS_comparison}
\end{figure*}

\subsection{DAPSs}

Based on the joint observation of the omni-directional PDP and the spatially resolved directional power angular spectrum (DAPS), several key frequency-dependent propagation characteristics can be identified. Notably, their absolute power levels are not directly comparable. This is because the DAPS is obtained via SAGE with antenna pattern compensation, which resolves multipath components in the angular domain, whereas the PDP represents the aggregated received energy and is inherently affected by antenna gain. 

Across both LOS and NLOS scenarios, a clear frequency-dependent transition in the effective channel representation can be observed. While the geometric structure of the propagation environment remains largely unchanged, the set of observable multipath components varies significantly with frequency. In the LOS scenario, strong spatial consistency is maintained across frequencies. The dominant propagation clusters—including the direct path at approximately 106~m and the primary reflections around 110–120~m—exhibit nearly identical delay and angular locations at both 8~GHz and 15~GHz, confirming that the underlying propagation geometry is primarily determined by the physical environment and is inherently frequency-independent. However, the richness of the multipath structure differs substantially. At 8~GHz, the DAPS reveals a dense distribution of diffuse scattering components across both delay and angular domains, whereas at 15~GHz, many of these weaker components are significantly attenuated and fall below the noise floor, resulting in a channel dominated by a limited number of strong specular paths. This frequency-dependent sparsification becomes more pronounced in the NLOS scenario. At 8~GHz, both PDP and DAPS capture multiple long-delay clusters (e.g., around 194~m and 240~m). In contrast, at 15~GHz, these higher-order components experience substantial attenuation and are rarely detectable, leading to a channel representation with only a few observable clusters. These observations indicate that, although the physical propagation environment remains unchanged, the set of observable multipath components is strongly frequency-dependent. At higher frequencies, increased free-space path loss, higher reflection losses, and reduced diffraction capability jointly weaken higher-order propagation paths. As a result, many diffuse and long-delay components fall below the measurement dynamic range and cannot be reliably detected. Consequently, the channel appears progressively sparser at higher frequencies, not due to a change in the propagation geometry, but due to the reduced observability of weaker multipath components.

\subsection{Path Loss}
Path loss is a key metric for evaluating coverage capability in wireless communication systems. At each measurement location, 100 snapshots were collected, and the path loss at each location can be calculated using the following formula:
\begin{equation}
PL = -10\log_{10}\left(\frac{1}{N}\sum_{n=1}^{N}|h_{n}(\tau)|^2\right),
\end{equation}
where $h_{n}(\tau)$ is the CIR at the $n$-th delay bin, and $N$ is the number of delay bins.

\begin{figure}[!ht]
\begin{subfigure}{\columnwidth}
    \centering
    \includegraphics[width=1\columnwidth]{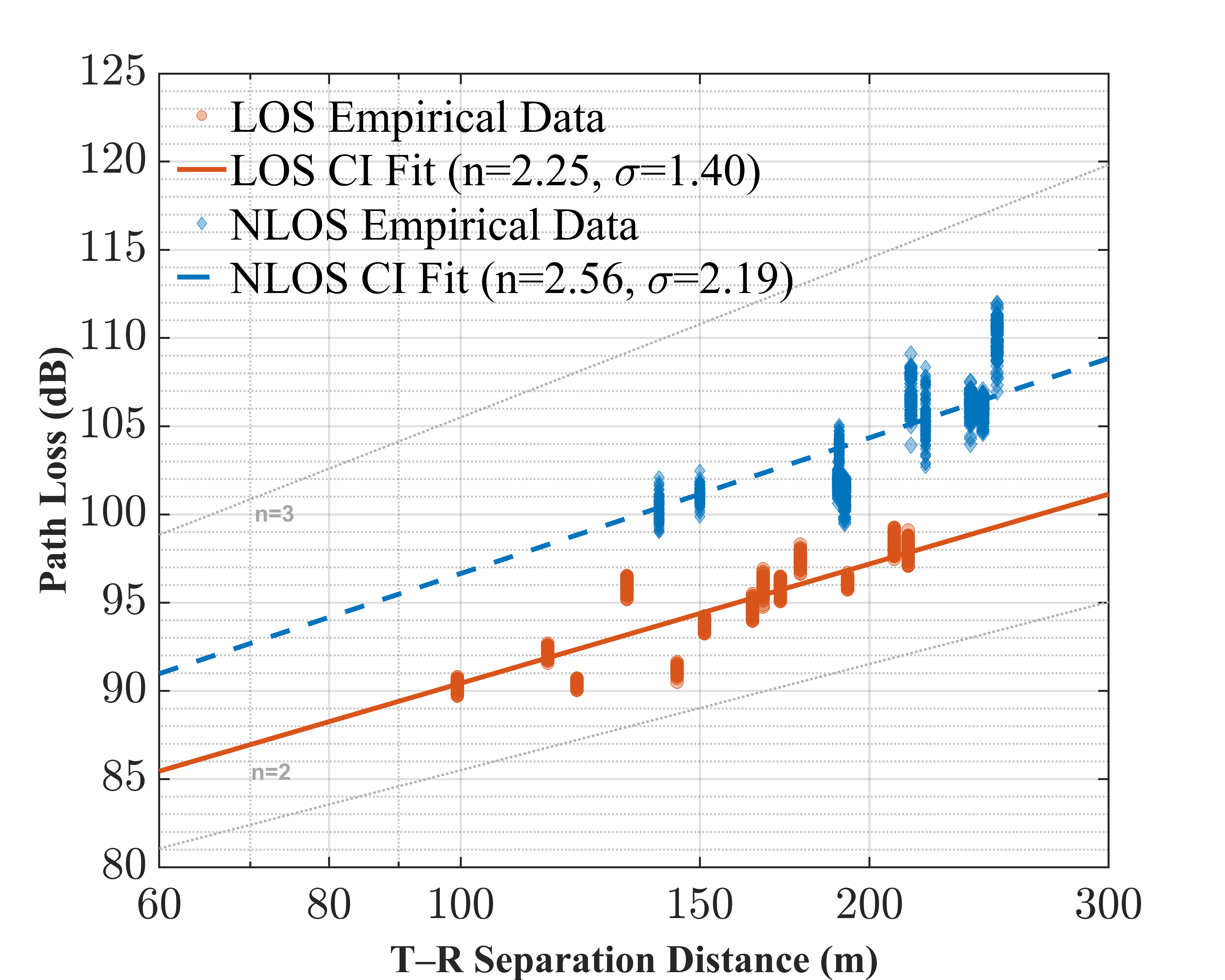}
    \caption{CI path loss model at 8 GHz}
    \label{fig:CI_8}
\end{subfigure}
\begin{subfigure}{\columnwidth}
    \centering
    \includegraphics[width=1\columnwidth]{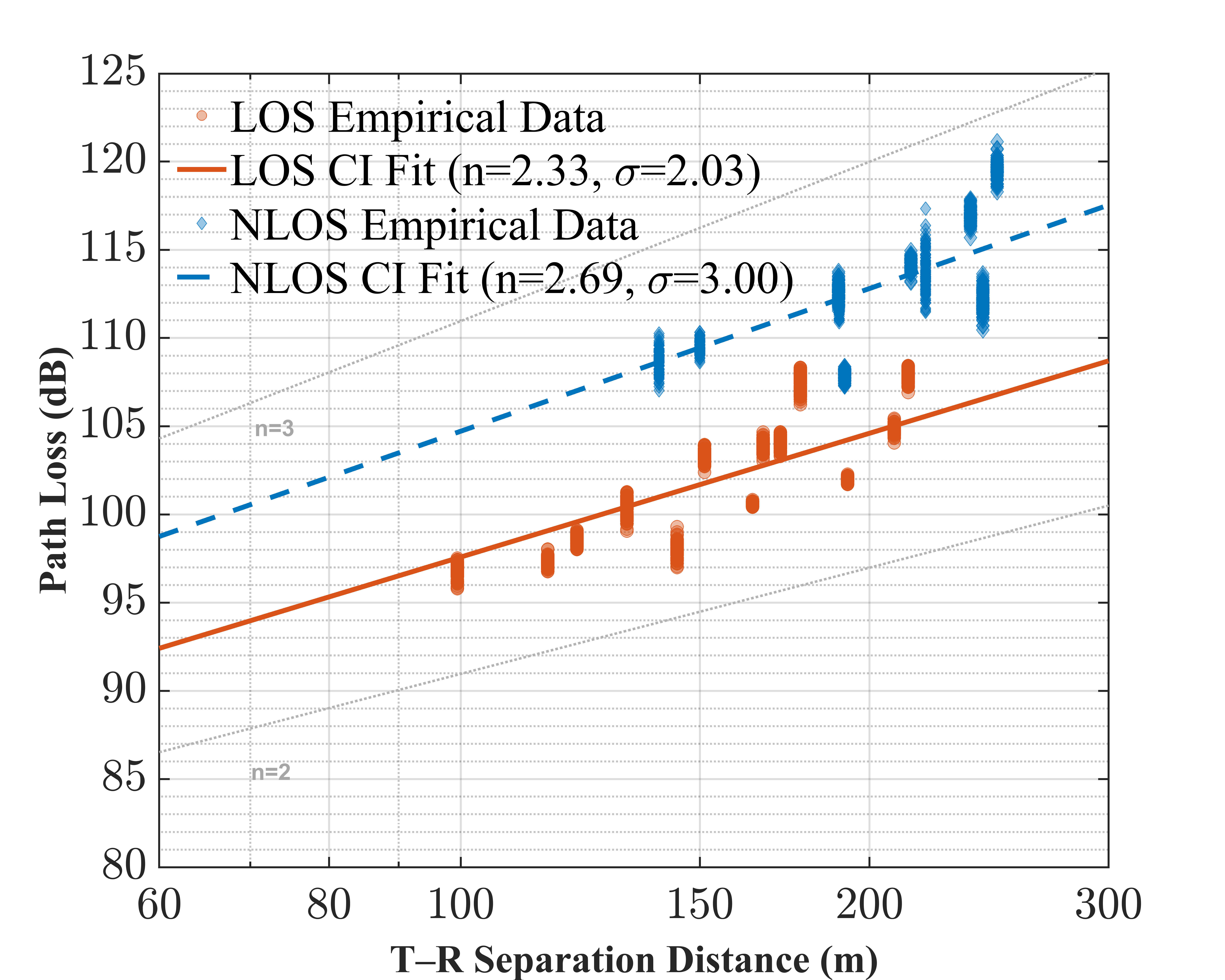}
    \caption{CI path loss model at 15 GHz}
    \label{fig:CI_15}
\end{subfigure}
\caption{CI path loss model fitted to measured data at 8 GHz and 15 GHz.}
\label{Fig.CI}
\end{figure}

Based on the obtained large-scale path loss values at different transmitter–receiver (T–R) separations, the close-in (CI) path loss model was employed to characterize the distance-dependent signal attenuation at 8~GHz and 15~GHz under both line-of-sight (LOS) and non-line-of-sight (NLOS) conditions \cite{CI}. 
\begin{equation}
PL_{CI}(d) = PL(d_0) + 10n\log_{10}\left(\frac{d_{3D}}{d_{0}}\right) + X_{\sigma},
\end{equation}
where $PL(d_0)$ is the free-space path loss (FSPL) in dB at a reference distance $d_0$, and is a function of wavelength $\lambda$ given by 
\begin{equation}
\label{equ.fspl}
PL(d_0) = 20\log_{10}\left(\frac{4\pi d_0}{\lambda}\right),
\end{equation}
with $d_0 = 1$~m. $n$ is the path-loss exponent (PLE). $X_{\sigma}$ is a zero-mean Gaussian random variable with a standard deviation $\sigma$ in dB, also known as the shadow factor, representing large-scale signal fluctuations resulting from shadowing by large obstructions in the wireless channel. We processed the measured data and fitted it using the CI path loss model. For comparison, reference curves with path loss exponents $n = 2$, and $3$ were also plotted, where $n = 2$ corresponds to the theoretical free-space path loss, as illustrated in Fig.~\ref{Fig.CI}.
The empirical path loss measurements and the corresponding CI model fits are generally consistent with established electromagnetic propagation principles. Overall, the absolute path loss at 15~GHz is higher than that at 8~GHz, which is primarily attributed to the frequency-dependent nature of free-space path loss.

Comparing the two frequency bands, both the path-loss exponent ($n$) and the shadow fading standard deviation ($\sigma$) exhibit a consistent increase when moving from 8~GHz to 15~GHz under both LOS and NLOS conditions. Specifically, the PLE increases from 2.25 to 2.33 in LOS and from 2.56 to 2.69 in NLOS, while the shadow fading standard deviation rises from 1.40~dB to 2.03~dB in LOS and from 2.19~dB to 3.00~dB in NLOS. This frequency-dependent behavior indicates that higher-frequency signals experience stronger large-scale attenuation and variability, even under the same propagation conditions. The increase in PLE can be attributed to the reduced diffraction capability and higher penetration loss at shorter wavelengths, which limit the ability of signals to propagate around obstacles. Meanwhile, the larger shadow fading standard deviation suggests that signal power becomes more sensitive to environmental features, leading to stronger spatial fluctuations. For completeness, comparing LOS and NLOS scenarios, both $n$ and $\sigma$ are consistently higher in NLOS conditions at each frequency band. This is expected, as NLOS propagation relies more on reflections, scattering, and diffraction, which introduce additional attenuation and variability. Overall, these results highlight that frequency plays a dominant role in large-scale fading characteristics, with higher frequencies imposing more stringent link budget requirements and greater sensitivity to environmental changes in UMa scenarios.

\subsection{RMS Delay Spread}
The RMS-DS is a fundamental parameter for characterizing the temporal dispersion of a wireless channel. It quantifies the spread of multipath components in the time domain and describes how signal energy is distributed around the mean arrival time.

\begin{figure}[!ht]
\centering
\begin{subfigure}{0.45\columnwidth}
    \centering
    \includegraphics[width=\textwidth]{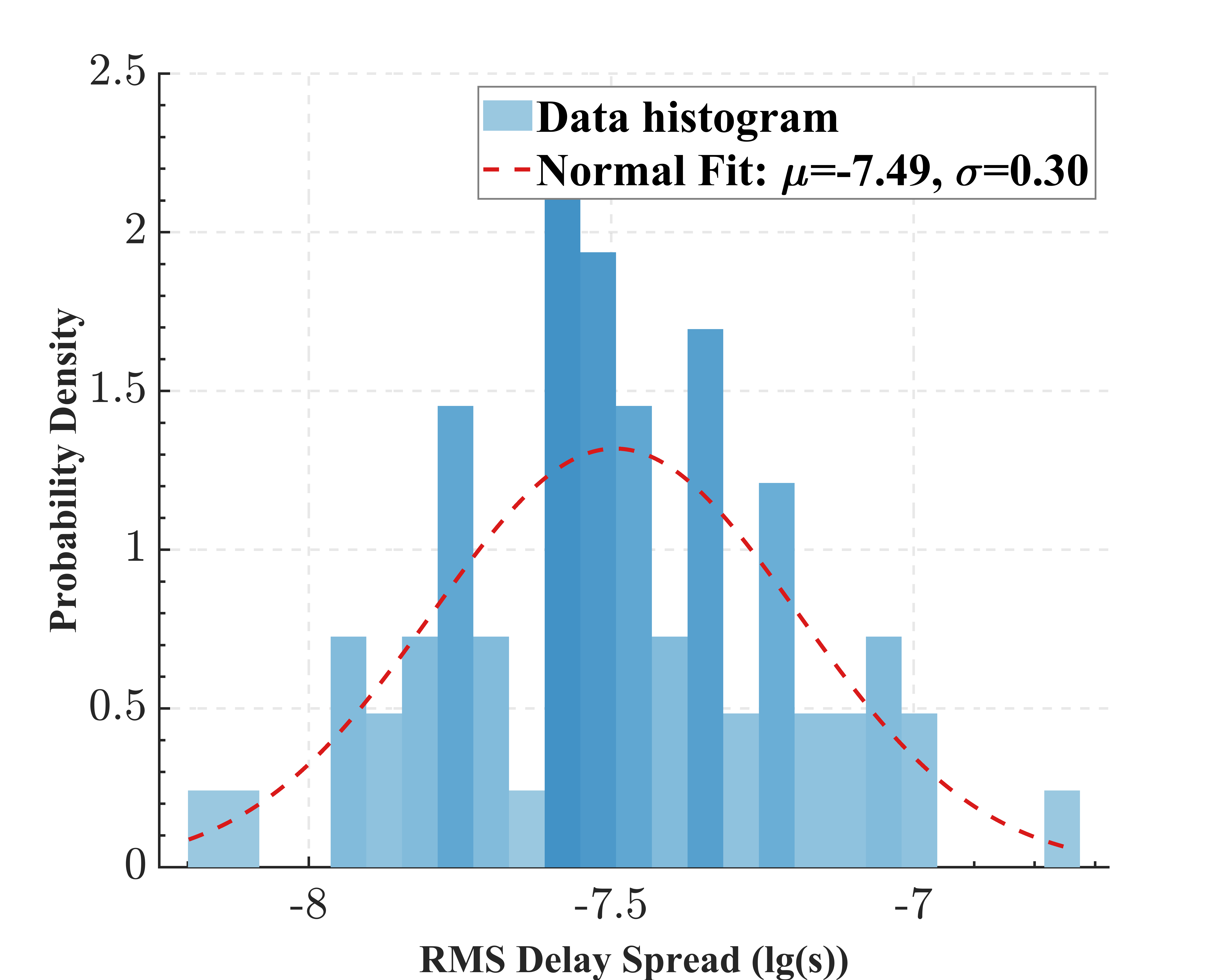}
    \caption{}
    \label{fig:los_8ghz_array}
\end{subfigure}
\hspace{0.05\columnwidth}
\begin{subfigure}{0.45\columnwidth}
    \centering
    \includegraphics[width=\textwidth]{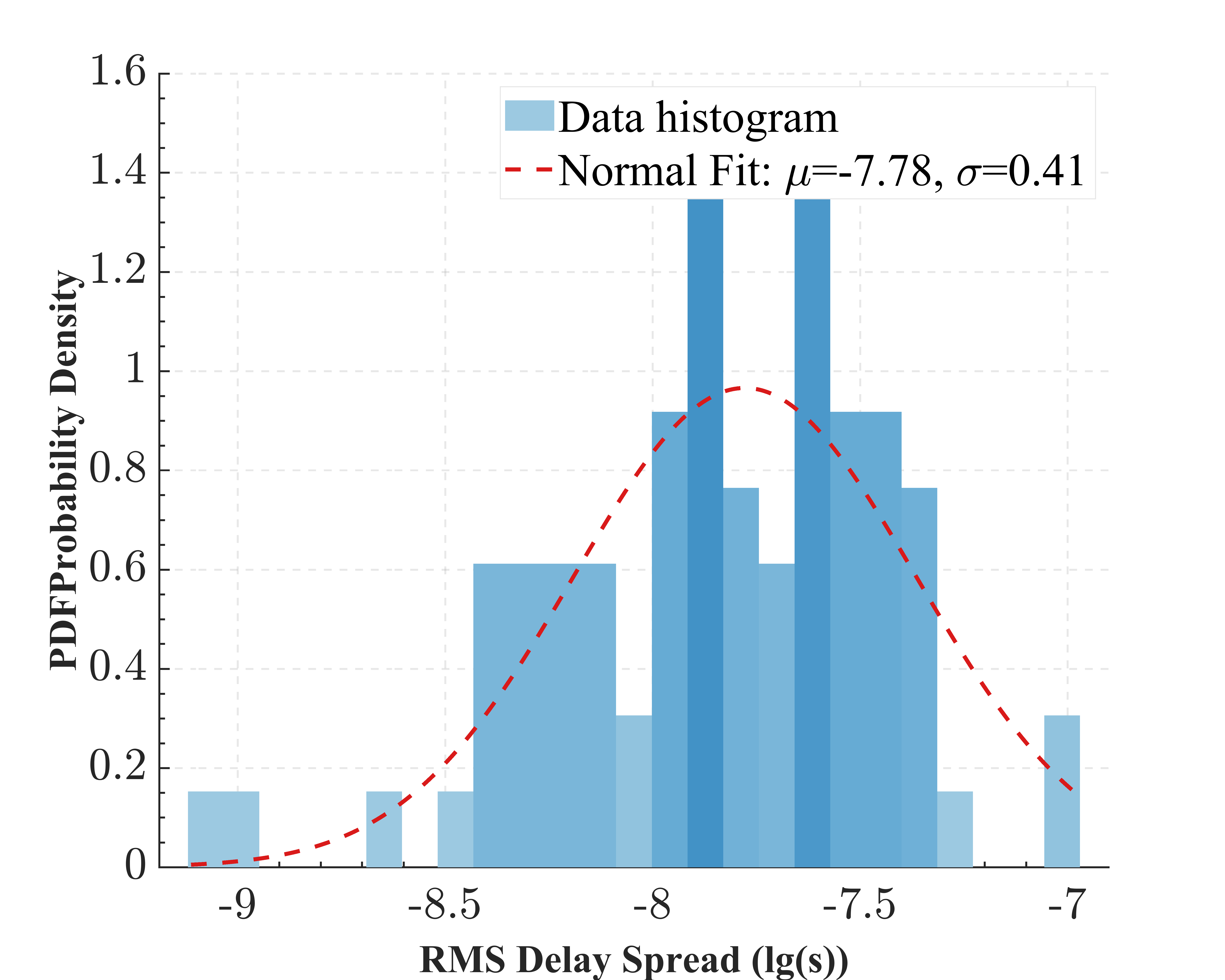}
    \caption{}
    \label{fig:los_15ghz_array}
\end{subfigure}
\\[0.5em]
\begin{subfigure}{0.45\columnwidth}
    \centering
    \includegraphics[width=\textwidth]{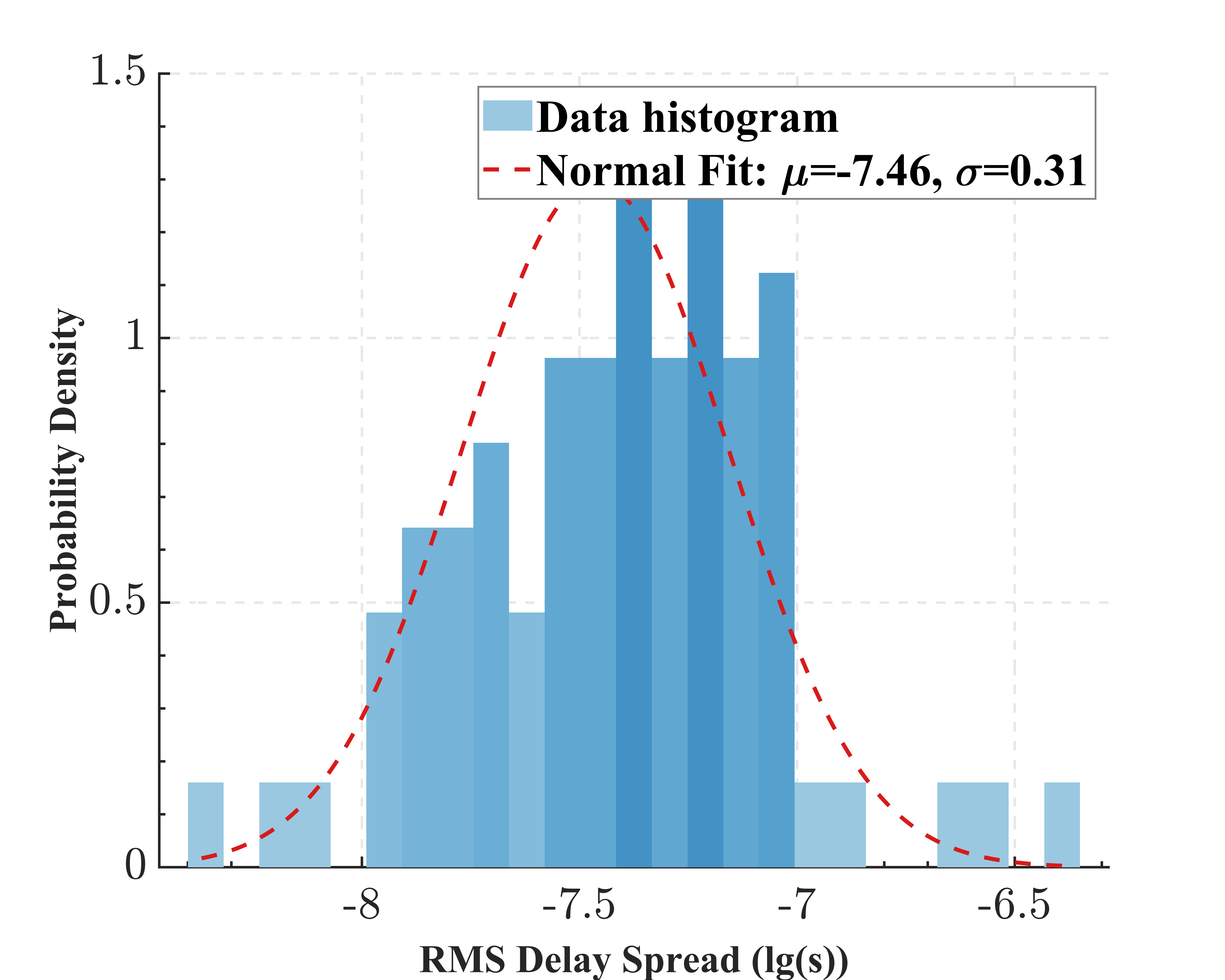}
    \caption{}
    \label{fig:nlos_8ghz_array}
\end{subfigure}
\hspace{0.05\columnwidth}
\begin{subfigure}{0.45\columnwidth}
    \centering
    \includegraphics[width=\textwidth]{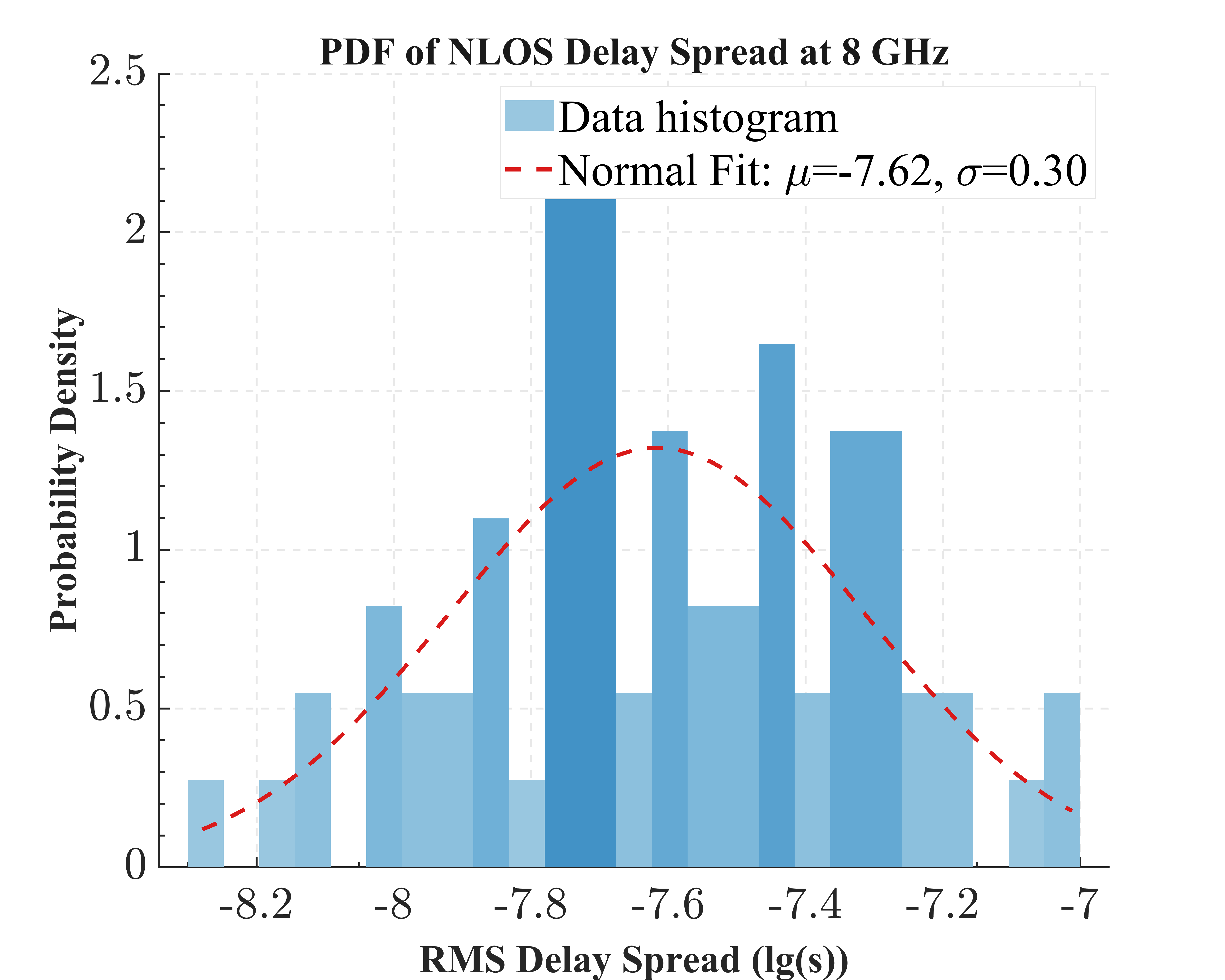}
    \caption{}
    \label{fig:nlos_15ghz_array}
\end{subfigure}
\caption{RMS delay spread results under same array aperture. 
(a) 8\,GHz LOS scenario; 
(b) 15\,GHz LOS scenario; 
(c) 8\,GHz NLOS scenario; 
(d) 15\,GHz NLOS scenario.}
\label{Fig.delay}
\end{figure}

Given the PDP $P(\tau_i)$ with delays $\tau_i$, the mean delay $\tau_\text{mean}$ and RMS Delay Spread $\tau_\text{rms}$ are defined as:
\begin{equation}
\tau_{mean} = \frac{\sum_{l=1}^L P(\tau_l)\tau_l}{\sum_{l=1}^L P(\tau_l)},
\end{equation}
\begin{equation}
\tau_{rms}=\sqrt{\frac{\sum_{l=1}^{L}{(\tau_{l}-\tau_{mean})^2}P(\tau_l)}{\sum_{l=1}^LP(\tau_l)}},
\end{equation}
where, $\tau_\text{mean}$ represents the average arrival time of the multipath components, while $\tau_\text{rms}$ indicates how these components are dispersed around the mean. A larger RMS-DS corresponds to stronger temporal dispersion, which may lead to inter-symbol interference (ISI). Conversely, a smaller RMS-DS suggests that multipath components are more temporally concentrated, resulting in a more stable received signal. Fig.~\ref{Fig.delay} and Table~\ref{tab:delayspread_samearea} summarize the RMS delay spread distributions for both 8~GHz and 15~GHz bands.
\begin{table}[bht!]
    \centering
    \caption{Statistical Parameters of RMS Delay Spread under same array aperture}
    \renewcommand{\arraystretch}{1.25} 
    \setlength{\tabcolsep}{10pt} 
    \begin{tabularx}{0.95\columnwidth}{l *{4}{>{\centering\arraybackslash}X}}
        \toprule
        \multirow{2}{*}{\textbf{Parameter}} & \multicolumn{2}{c}{\textbf{8 GHz}} & \multicolumn{2}{c}{\textbf{15 GHz}} \\
        \cmidrule(lr){2-3} \cmidrule(lr){4-5}
        & \textbf{LOS} & \textbf{NLOS} & \textbf{LOS} & \textbf{NLOS} \\
        \midrule
        $\mu$     & $-7.49$ & $-7.46$ & $-7.78$ & $-7.62$ \\
        $\sigma$  & $0.30$ & $0.31$ & $0.41$ & $0.30$ \\
        \bottomrule
    \end{tabularx}
    \label{tab:delayspread_samearea}
\end{table}

A clear frequency-dependent reduction in RMS delay spread is observed across both LOS and NLOS scenarios, indicating that the channel becomes more temporally compact at higher frequencies. Specifically, the mean value decreases from $\mu=-7.49$ to $\mu=-7.78$ in LOS and from $\mu=-7.46$ to $\mu=-7.62$ in NLOS when moving from 8~GHz to 15~GHz. This behavior reflects the reduced contribution of late-arriving multipath components at higher frequencies. Due to stronger attenuation, higher penetration loss, and diminished diffraction capability at shorter wavelengths, weaker and longer-delay paths are more likely to be suppressed below the noise floor. Consequently, the received energy becomes increasingly concentrated on a small number of dominant paths, leading to a more compact delay-domain representation. The variability of the delay spread further reveals scenario-dependent characteristics. In the LOS scenario, the standard deviation increases from $\sigma=0.30$ at 8~GHz to $\sigma=0.41$ at 15~GHz, suggesting enhanced sensitivity to local environmental changes. This can be attributed to the intermittent presence or blockage of dominant specular paths, which has a more pronounced impact when the number of effective paths is limited. In contrast, in the NLOS scenario, the standard deviation remains nearly unchanged (from $\sigma=0.31$ to $\sigma=0.30$), while the mean delay spread decreases. This indicates that the delay spread is not only reduced but also confined within a narrower range at higher frequencies. Such behavior suggests that strong attenuation consistently suppresses higher-order multipath components across different locations, resulting in a more uniformly compact channel. Overall, these results demonstrate that increasing frequency leads to a systematically more compact and constrained delay-domain channel representation, driven by the reduced observability of weak and late-arriving multipath components.

\subsection{RMS Angle Spread}

\begin{figure*}[t]
    \centering
    \begin{subfigure}[b]{0.48\textwidth}
        \includegraphics[width=\textwidth]{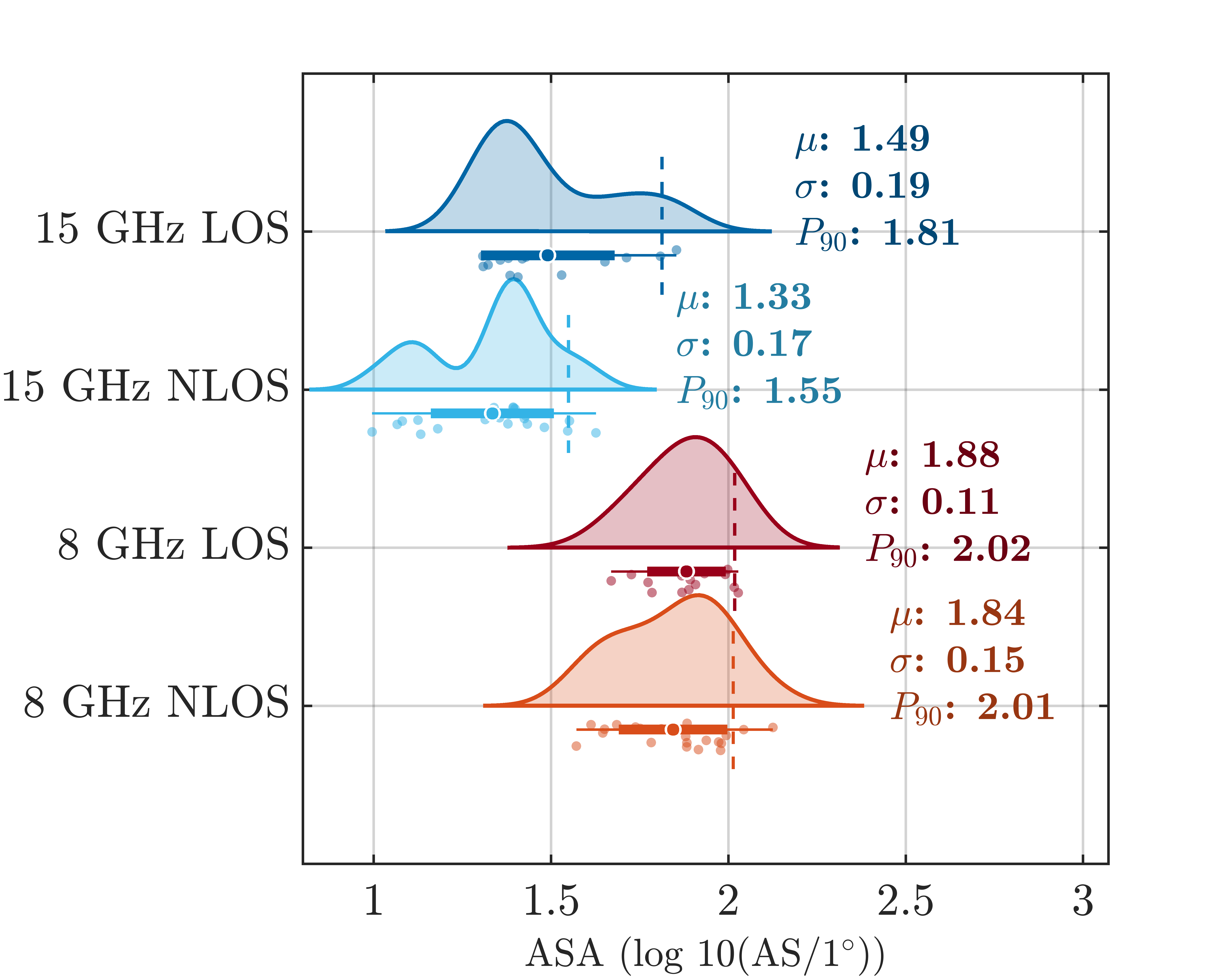}
        \caption{RMS ASA}
        \label{fig:asa_raincloud}
    \end{subfigure}
    \hfill
    \begin{subfigure}[b]{0.48\textwidth}
        \includegraphics[width=\textwidth]{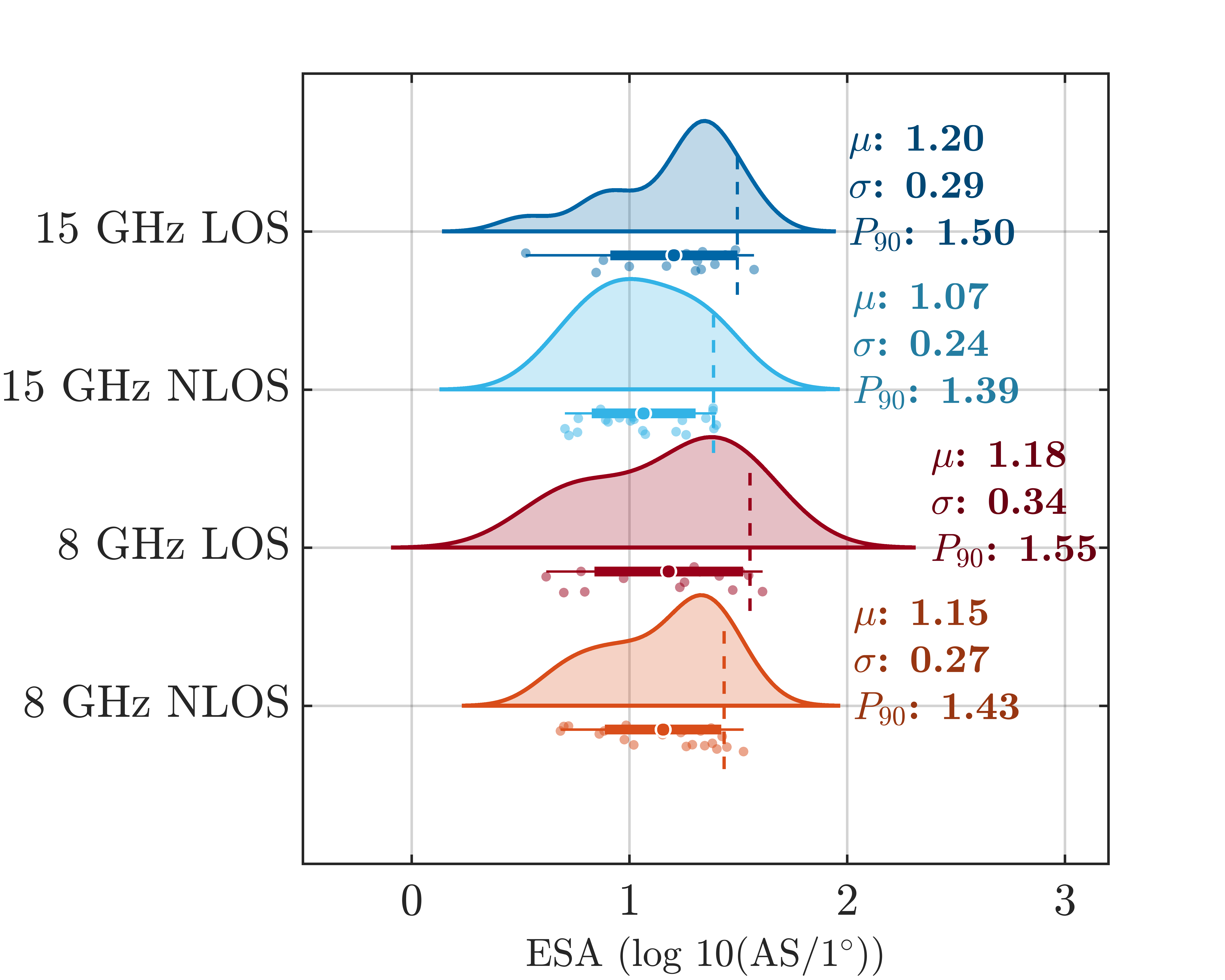}
        \caption{RMS ESA}
        \label{fig:esa_raincloud}
    \end{subfigure}
    \caption{Raincloud plots of the measured RMS angular spreads for 8 GHz and 15 GHz bands under equal physical aperture constraints. }
    \label{fig:Angular_Spread_Raincloud}
\end{figure*}

While RMS Delay Spread captures the temporal dispersion of multipath components, the RMS-AS characterizes their spatial dispersion. Quantifies how the power of the multipath components is distributed around the mean arrival or departure angle, thus describing the angular richness of the channel. For multipath components with angles $\theta_i$ and powers $P(\theta_i)$, the mean angle $\theta_\text{mean}$ and RMS Angle Spread $\theta_\text{rms}$ are expressed as \cite{zhang}:
\begin{equation}
\theta_\text{mean} = \frac{\sum_{i=1}^{N} P(\theta_i) \theta_i}{\sum_{i=1}^{N} P(\theta_i)},
\end{equation}
\begin{equation}
\theta_\text{rms} = \sqrt{\frac{\sum_{i=1}^{N} P(\theta_i) (\theta_i - \theta_\text{mean})^2}{\sum_{i=1}^{N} P(\theta_i)}},
\end{equation}
where, $\theta_\text{mean}$ denotes the average angle of arrival or departure, and $\theta_\text{rms}$ represents the angular spread around this mean. A larger RMS-AS implies that multipath components arrive from a wider range of directions. In contrast, a smaller RMS-AS indicates that the energy received is concentrated within a narrow angular region, facilitating directional transmission and improving array efficiency. The ASA and ESA is shown in Fig.\ref{fig:Angular_Spread_Raincloud}.

\begin{table}[t]
\centering
\caption{Statistical Parameter of RMS Angular Spread under same array aperture}
\label{tab:as_stats}
\renewcommand{\arraystretch}{1.2}
\setlength{\tabcolsep}{6pt}

\begin{tabular*}{\columnwidth}{@{\extracolsep{\fill}}lcccc}
\toprule
\textbf{Scenario} 
& \multicolumn{2}{c}{\textbf{ASA (log$_{10}$(AS/$1^{\circ}$))}} 
& \multicolumn{2}{c}{\textbf{ESA (log$_{10}$(AS/$1^{\circ}$))}} \\
\cmidrule(lr){2-3} \cmidrule(lr){4-5}
& $\boldsymbol{\mu}$ & $\boldsymbol{\sigma}$ 
& $\boldsymbol{\mu}$ & $\boldsymbol{\sigma}$ \\
\midrule
15\,GHz LOS  & 1.49 & 0.19 & 1.20 & 0.29 \\
15\,GHz NLOS & 1.33 & 0.17 & 1.07 & 0.24 \\
8\,GHz LOS   & 1.88 & 0.11 & 1.18 & 0.34 \\
8\,GHz NLOS  & 1.84 & 0.15 & 1.15 & 0.27 \\
\bottomrule
\end{tabular*}
\end{table}
A pronounced reduction in angular dispersion is observed at higher frequencies, which is predominantly manifested in the azimuth domain. At 15GHz, the mean log-domain ASA ($\mu$) is consistently and significantly smaller than that at 8GHz across both LOS and NLOS scenarios, accompanied by a substantial decrease in the upper percentile ($P_{90}$). This indicates that the multipath energy becomes highly concentrated within a narrower angular sector. Such behavior is consistent with the sparse multipath structures observed in the DAPS (Fig.\ref{fig:DPAS_comparison}), where only a limited number of dominant directions remain at higher frequencies. In contrast, the variation of ESA across frequencies is considerably less pronounced and exhibits a clear scenario dependency. Under LOS conditions, the ESA remains largely unchanged when scaling from 8GHz to 15GHz, indicating that the vertical angular distribution is primarily constrained by the geometric layout of the environment. In NLOS scenarios, a moderate reduction in ESA is observed at 15~GHz; however, this decrease is significantly smaller than that observed in the azimuth domain. This distinction suggests that the frequency-induced contraction of angular dispersion is inherently anisotropic. While the increased propagation loss at higher frequencies reduces the overall multipath richness, its impact is more pronounced in the azimuth domain, where scattering is inherently richer. In contrast, the elevation domain, being geometrically constrained and less diverse, exhibits only a limited sensitivity to frequency scaling.

\section{Coverage Performance Analysis under Equal Physical Aperture Constraints}
\subsection{Theoretical Coverage Bound}

With the arrays constructed to achieve approximately identical physical apertures at 8~GHz and 15~GHz, we now analyze the measured coverage performance under the constraint of equal total transmit power $P_t$ and fixed array aperture $A_{\rm array}$. The received power $P_r$ in free space is described by the Friis transmission equation:
\begin{equation}
P_r = P_t G_t G_r \left( \frac{\lambda}{4 \pi d_{\rm prop}} \right)^2,
\label{equ.friis_correct}
\end{equation}
where $G_t$ and $G_r$ are the transmit and receive array gains, $\lambda$ is the wavelength, and $d_{\rm prop}$ is the propagation distance. For a UPA with half-wavelength element spacing $d = \lambda/2$, the transmit array gain scales linearly with the number of elements $N$:
\begin{equation}
G_t \approx N \cdot G_{\rm element}, \quad N = \frac{A_{\rm array}}{(\lambda/2)^2}.
\label{equ.array_gain}
\end{equation}
Substituting Eq.~(\ref{equ.array_gain}) into Eq.~(\ref{equ.friis_correct}), we have
\begin{equation}
P_r \propto P_t \cdot \frac{A_{\rm array}}{\lambda^2} \cdot G_r \cdot \left( \frac{\lambda}{4 \pi d_{\rm prop}} \right)^2 
= P_t G_r \frac{A_{\rm array}}{(4 \pi d_{\rm prop})^2}.
\end{equation}

This derivation clearly shows that under the conditions of fixed total transmit power and fixed physical aperture with half-wavelength spacing, the received power $P_r$ becomes independent of the carrier frequency. This result indicates that when the total transmit power and physical array aperture are fixed, increasing the number of antenna elements at higher frequencies compensates for the increased free-space path loss through coherent array gain. Consequently, the achievable received power remains approximately invariant with respect to the carrier frequency. It is crucial to note that this theoretical frequency invariance heavily relies on the assumption of ideal coherent spatial processing, where the phases of all $N$ antenna elements are perfectly aligned.

\begin{figure}[!ht]
\centering
\includegraphics[width=1.0\columnwidth]{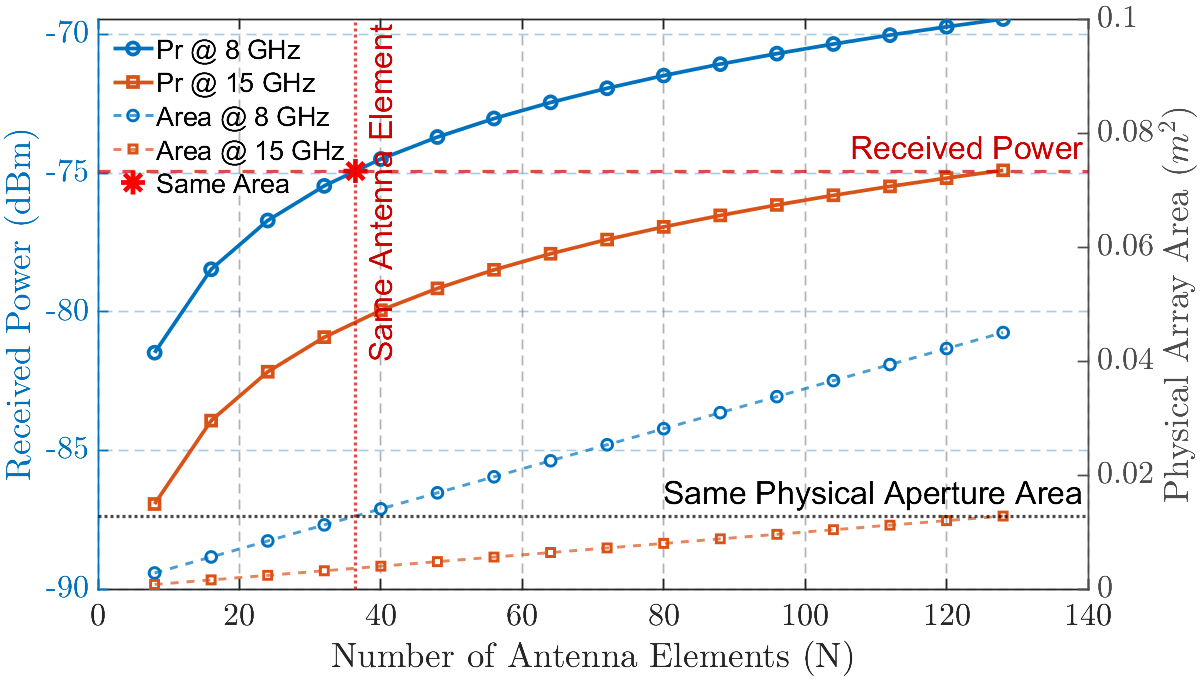}
\caption{Numerical relationship of theoretical received power under fixed-aperture and half-wavelength spacing conditions. }
\label{Fig.Pr_area}
\end{figure}
Fig.~\ref{Fig.Pr_area} numerically validates this principle. The 128-element 15~GHz array occupies nearly the same physical area as the 32-element 8~GHz array, resulting in almost identical received power levels despite the difference in wavelength. This provides a solid baseline for subsequent coverage and spectral efficiency analyses.

\subsection{Non-Coherent Combining Gain}
\begin{figure}[!ht]
    \centering
    \includegraphics[width=1\columnwidth]{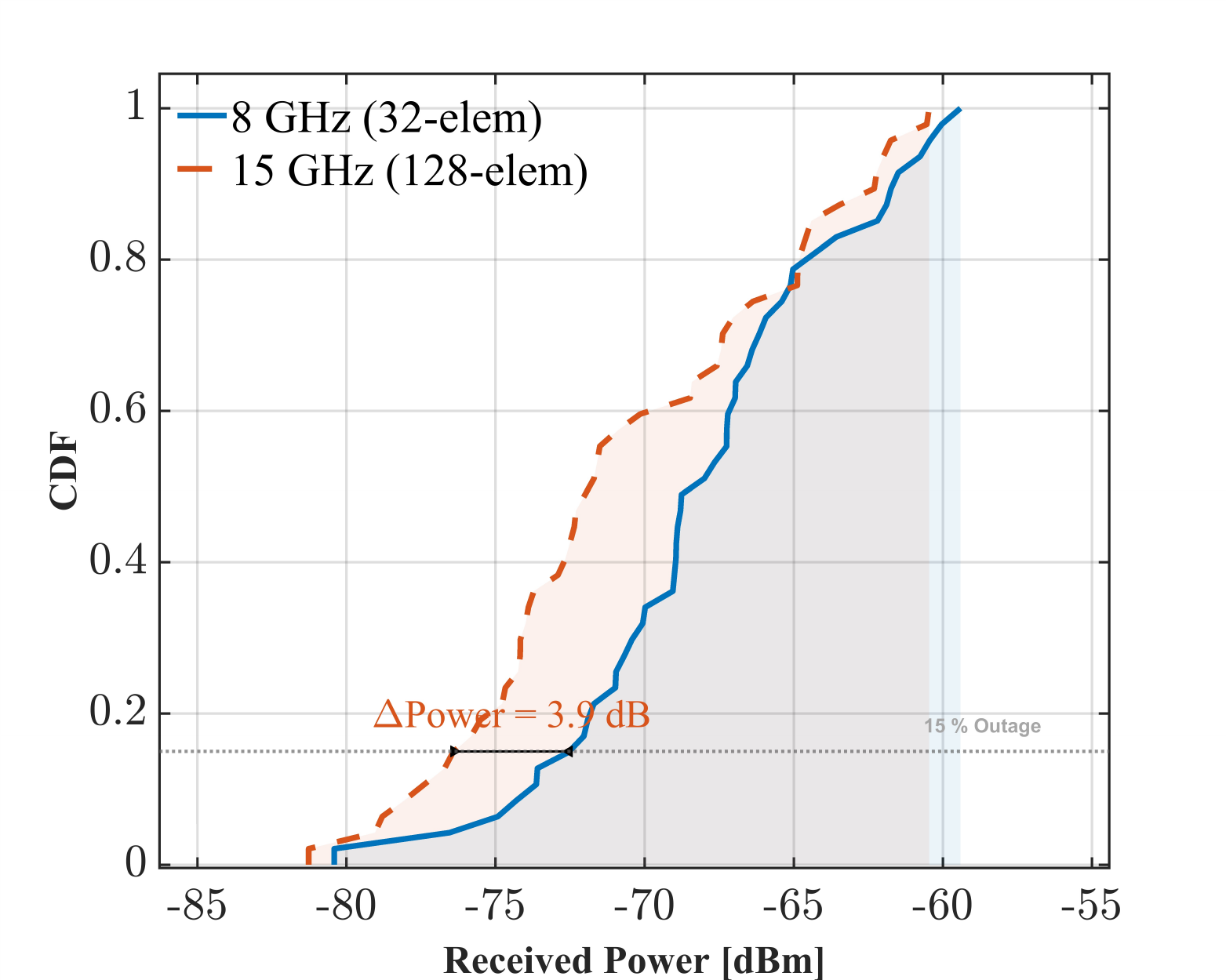} 
    \caption{CDF of the receive power (non-coherent combining), comparing 8 GHz (32 elements) with 15 GHz (128 elements).}
    \label{Fig.Pr_NMC}
\end{figure}

While the theoretical analysis in Section A establishes that frequency-independent coverage relies on ideal coherent combining, this assumption may not hold in practical propagation environments. To explicitly reveal the impact of the channel itself, we first analyze the non-coherent received power as a baseline. This metric quantifies the total raw electromagnetic energy captured by the array, independent of phase alignment, thereby exposing the intrinsic propagation differences—such as multipath sparsity and scattering losses—between the two frequency bands. This section quantifies the received power disparity between the 8 GHz system ($N_{8}=32$) and the 15 GHz system ($N_{15}=128$). The total non-coherent received power, $P_{nc}$, is defined as the summation of the energy integrated across all transmit-receive antenna pairs: 

\begin{equation}
P_{nc} = \sum_{t=1}^{N_{tx}} \sum_{r=1}^{N_{rx}} \sum_{d=1}^{N_d} |h_{d,t,r}|^2,
\label{eq:Pnc}
\end{equation}
where $h_{d,t,r}$ represents the empirical channel impulse response (CIR) derived from measurements. To ensure a fair comparison under the constant physical aperture constraint, the total transmit power is held constant across both frequency configurations ($P_{tot, 8\text{GHz}} = P_{tot, 15\text{GHz}}$). The empirical Cumulative Distribution Function (CDF) of the non-coherent received power, derived from field measurements in UMa environments, is presented in Fig. \ref{Fig.Pr_NMC}.

As shown in the figure, several key observations regarding high-frequency coverage can be made. At the $15\%$ outage probability level (0.15 CDF), the 15 GHz system exhibits a residual coverage deficit of approximately 3.9 dB relative to the 8 GHz baseline. This indicates that, in practical urban propagation environments, the theoretical aperture-based compensation ($N \propto 1/\lambda^2$) does not fully offset the increased penetration loss and environmental variability at 15 GHz. Additionally, the 15 GHz CDF curve exhibits a steeper slope in the low-power region compared to 8 GHz, indicating a more concentrated distribution of the received power. This behavior is consistent with the increased channel sparsity discussed in Section III. At 15 GHz, the channel energy is dominated by a limited number of strong paths, resulting in less variation among most samples and thus a steeper CDF. In contrast, the richer diffuse multipath components at 8 GHz introduce greater variability in the received power, leading to a more spread-out distribution and a flatter CDF. The observed 3.9 dB gap further highlights the limitations of non-coherent energy combining in sparse high-frequency channels. Although the 128-element array at 15 GHz provides a significantly larger number of antennas, the use of non-coherent power summation prevents the system from fully exploiting the potential array gain. In sparse FR3 channels, where the received energy is concentrated in a few dominant propagation paths, performance critically depends on the ability to coherently combine these components. In contrast, the 8 GHz channel benefits from richer multipath propagation, where multiple components contribute to the received signal even without phase alignment, resulting in more robust power accumulation. Therefore, the observed performance gap indicates that simply increasing the number of antennas is insufficient to compensate for frequency-dependent propagation loss.

\subsection{Coherent Beamforming Gain}

To evaluate the ultimate coverage limit, we analyze the coherent beamforming performance using Dominant Eigenmode Transmission (DET) \cite{MRC}. By aligning the transceiver weights with the strongest spatial path, the 15~GHz system can theoretically provide up to 6~dB additional array gain under ideal coherent combining conditions. This massive spatial focusing capability is the primary mechanism intended to compensate for the severe free-space path loss and penetration attenuation inherent to the FR3 band.

\begin{figure}[!ht]
\centering
\includegraphics[width=1\columnwidth]{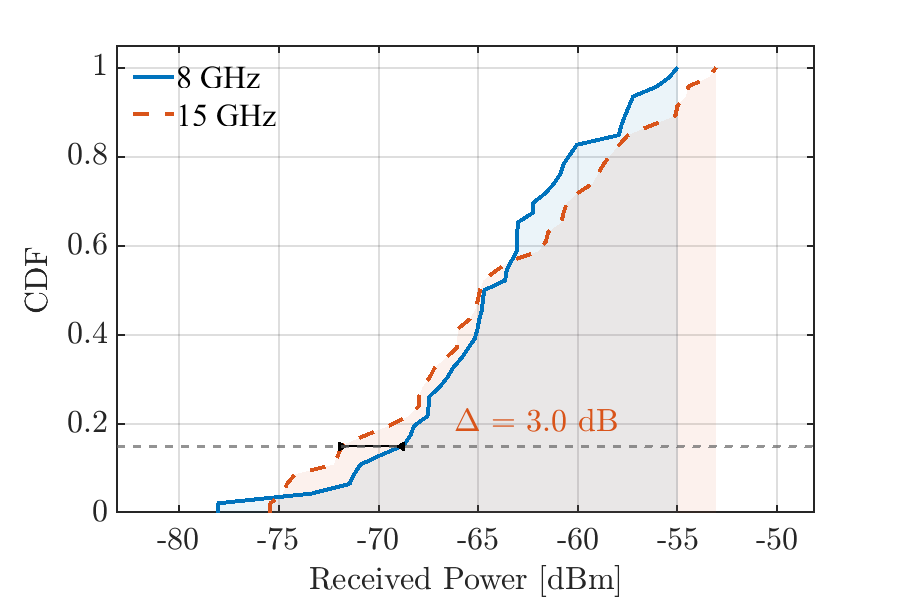} 
\caption{CDF of received power (optimal coherent beamforming) comparing 8~GHz (32 elements) and 15~GHz (128 elements).}
\label{Fig.CDF_MRC}
\end{figure}

As illustrated in Fig.~\ref{Fig.CDF_MRC}, the application of coherent beamforming reveals a scenario-dependent coverage relationship between the two frequency bands. In the low-power regime, corresponding to the $15\%$ outage threshold, the 15~GHz system remains approximately 3.0~dB below the 8~GHz baseline. This residual gap suggests that, under severe blockage conditions, the array gain provided by the increased number of antenna elements is insufficient to fully compensate for the additional diffraction and penetration losses at higher frequencies. In contrast, as the received power increases, a crossover behavior is observed. Beyond approximately $-58$~dBm, the 15~GHz system begins to outperform the 8~GHz baseline. This transition reflects the fact that high-power conditions are typically dominated by a small number of strong propagation paths. In such cases, coherent beamforming can effectively align with these dominant components, allowing the 15~GHz system to fully exploit its larger array size and achieve higher array gain. However, in the low-power regime, the performance improvement remains limited. Under these conditions, the channel is often characterized by severe blockage or the absence of strong dominant paths. As a result, even optimal coherent combining cannot provide substantial gain, since there is insufficient signal energy to be constructively combined. This explains why a residual performance gap persists at the cell edge despite the use of coherent beamforming. Overall, these results demonstrate that, under a fixed physical aperture constraint, the coverage performance of the 15~GHz system is inherently condition-dependent. While coherent beamforming is highly effective in scenarios with dominant propagation paths, its ability to compensate for coverage loss is fundamentally constrained in deeply shadowed environments.

\subsection{Array Topology Optimization}

\begin{figure}[htbp]
    \centering
    \includegraphics[width=0.95\columnwidth]{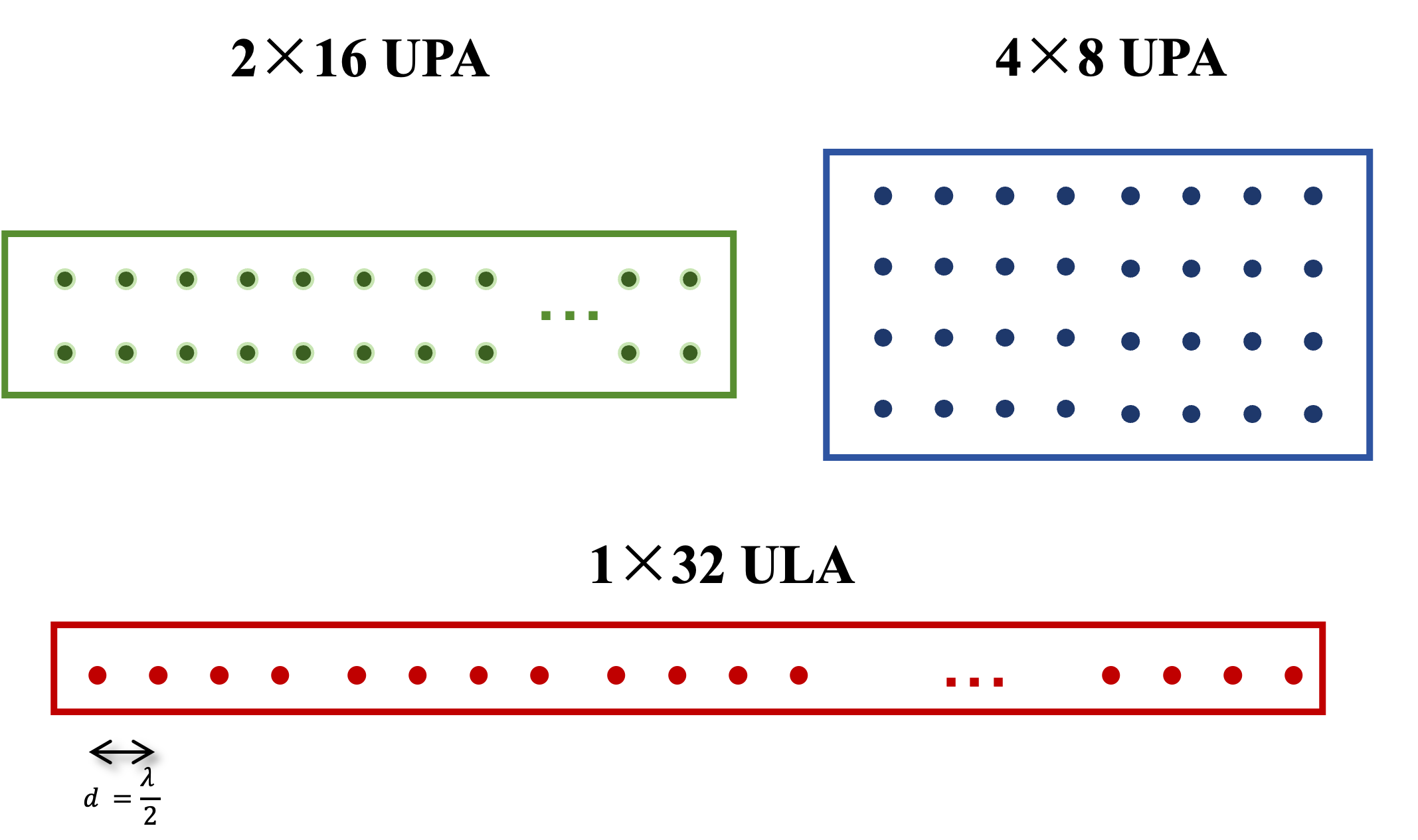}
    \caption{Illustration of array topologies reconstructed from the $4 \times 32$ measurement grid. Horizontal aperture increases from $4\times8$ to $1\times32$, improving azimuthal resolution.}
    \label{Fig.Array_Topology}
\end{figure}

To further enhance system performance under practical deployment conditions, the impact of antenna array topology on coverage capability is investigated by reconfiguring the element allocation within a fixed physical aperture. A $4 \times 32$ dual-polarized measurement grid at 8~GHz serves as the baseline from which several 32-element sub-array configurations are reconstructed: 
\begin{itemize}
    \item $1 \times 32$ ULA: all elements aligned horizontally, maximizing azimuthal aperture;
    \item $2 \times 16$ ULA: moderate horizontal and vertical extension, representing a compromise between azimuth and elevation resolution;
    \item $4 \times 8$ UPA: a more square-like configuration, emphasizing vertical resolution.
\end{itemize}
As illustrated in Fig.~\ref{Fig.Array_Topology}, increasing the horizontal aperture from the $4 \times 8$ to the $1 \times 32$ configuration results in a progressively narrower azimuthal beamwidth and higher angular resolution. This spatial reconfiguration aligns the array's sensing capability with the dominant propagation characteristics of the UMa environment, where horizontal scattering is inherently richer than vertical scattering.

\begin{figure}[htbp]
    \centering
    \includegraphics[width=1\columnwidth]{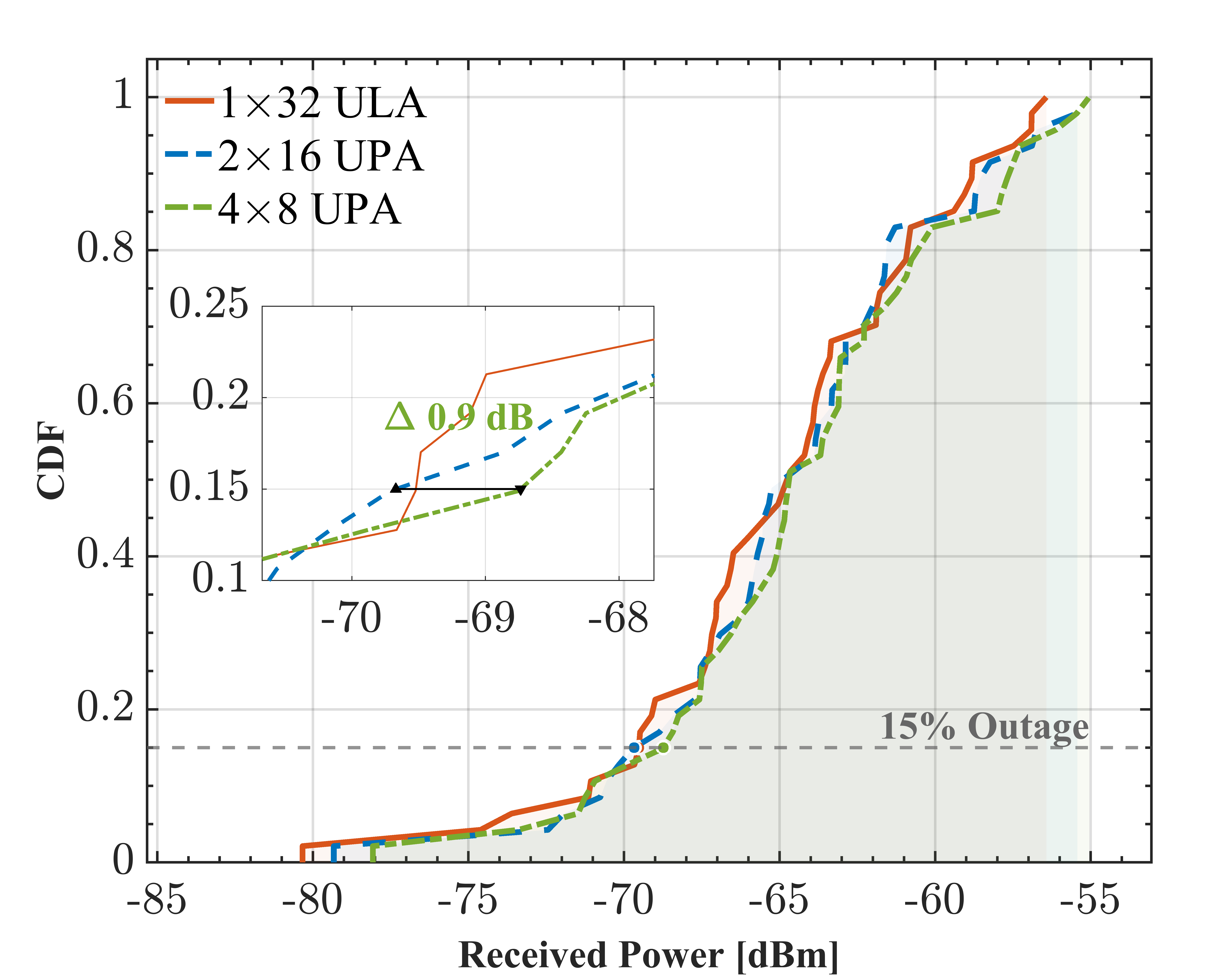} 
    \caption{Empirical CDF of received power for different array topologies at 8 GHz under optimal coherent beamforming.}
    \label{Fig.RxPower_CDF}
\end{figure}

The empirical CDF of the synthesized received power for the three array topologies is shown in Fig.~\ref{Fig.RxPower_CDF}. It can be observed that the distributions of the $1 \times 32$, $2 \times 16$, and $4 \times 8$ configurations almost completely overlap over the majority of the probability range. This indicates that the total received power is primarily determined by the aggregate array gain, which is fixed by the total number of antenna elements ($N=32$), rather than the specific geometric arrangement of the array. As a result, redistributing the elements across different topologies has a negligible impact on the overall captured energy. A slight deviation is only observed in the low-power regime corresponding to cell-edge conditions. At the 15\% outage level, the $1 \times 32$ ULA exhibits a marginal gain of approximately 0.9~dB compared to the $4 \times 8$ UPA. This difference can be attributed to small variations in directional sensitivity across array geometries under severe fading conditions. However, given that the magnitude of this gain is well below 1~dB, its impact on practical system performance is minimal. Overall, the results demonstrate that the received power is highly insensitive to array topology, implying that compact planar arrays can be adopted without compromising coverage performance.

\section{Spectral Efficiency Analysis under Equal Physical Aperture Constraints}
\subsection{Theoretical Spectral Efficiency Bound}

To establish a fundamental capacity upper bound and evaluate the intrinsic potential of the FR3 channel, we analyze the theoretical SE under an open-loop transmission framework assuming equal power allocation \cite{capacity}:

\begin{equation}
SE = \log_2 \det \left( \mathbf{I}_{N_r} + \frac{\rho}{N_t} \mathbf{H}\mathbf{H}^H \right),\label{equ.se_theory}
\end{equation}
where $\mathbf{H} \in \mathbb{C}^{N_r \times N_t}$ represents the channel matrix, and $\rho$ denotes the reference receive SNR. In this theoretical framework, we deliberately do not consider specific beamforming strategies. By fixing $\rho$ at 20~dB across all configurations and assuming an idealized uncorrelated full-rank channel, we remove variations caused by differences in link budget, beamforming design, synchronization errors, or hardware imperfections. This approach allows us to focus purely on the impact of the characteristics of channel.

\begin{figure}[!ht]
\centering
\includegraphics[width=1\columnwidth]{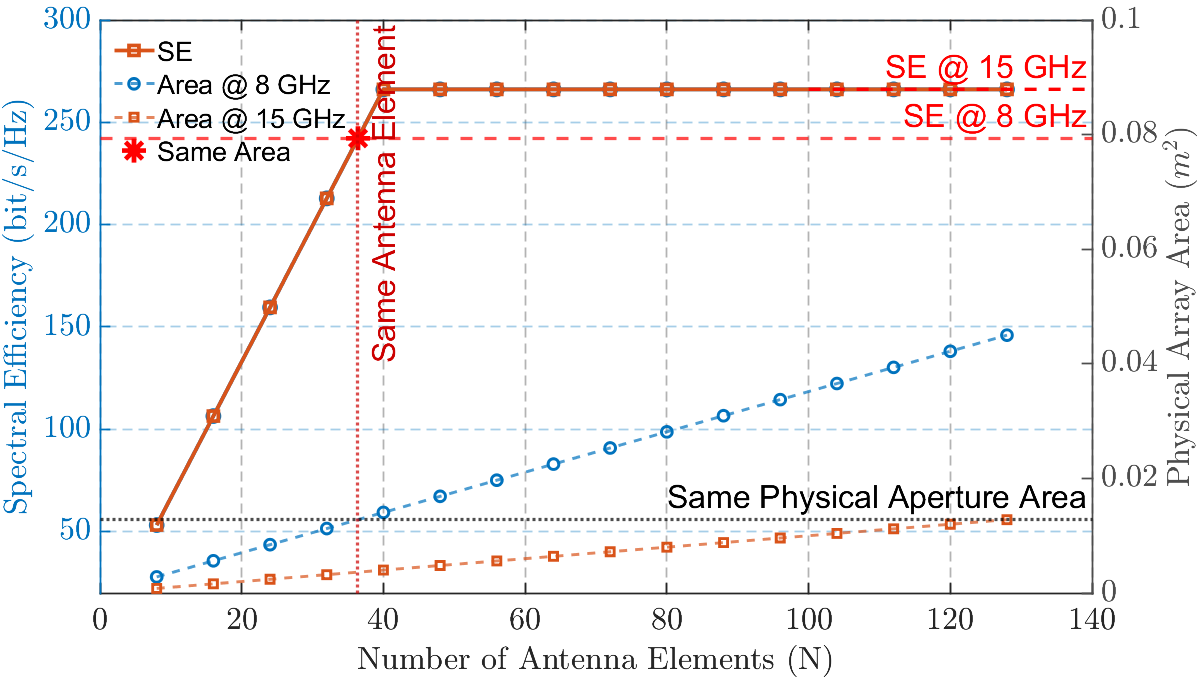}
\caption{Numerical relationship of theoretical received power under fixed-aperture and half-wavelength spacing conditions.}
\label{Fig.SE_theory}
\end{figure}

As shown in Fig.~\ref{Fig.SE_theory}, the SE scaling exhibits two operational regimes determined by the receiver size ($N_r = 40$): Multiplexing-Limited Regime ($N_t \le N_r$): In this region, SE increases approximately linearly with the number of transmit antennas, as each additional element adds spatial DoF and enables more independent data streams;Saturation Regime ($N_t > N_r$): When the number of transmit antennas exceeds the receiver rank, the available spatial DoF is fully utilized. Further increases in $N_t$ only provide modest improvements through array gain, while the multiplexing contribution is capped. A critical comparison is performed at the aperture-matched point (indicated by the red star in Fig.~\ref{Fig.SE_theory}). To perfectly match the physical area of the 15~GHz 128-element array ($0.0128~\text{m}^2$), the 8~GHz system would theoretically require $N \approx 36.4$ elements. However, in our practical measurement baseline, a standard 32-element configuration is adopted. At $N=32$, the 8~GHz system remains constrained in the multiplexing-limited regime, failing to fully occupy the available 40 spatial dimensions of the receiver. In contrast, the 15~GHz system, by accommodating 128 elements within the same footprint, fully saturates the spatial potential of the link.This comparison demonstrates that the 15~GHz 128 array elements system provides a significantly higher capacity ceiling (266.11~bit/s/Hz) than the practical 8~GHz 32 elements system baseline (242.20~bit/s/Hz). This sets the stage for the empirical analysis.

\subsection{Experimental Validation of Spectral Efficiency}

To assess the practical performance of FR3 Massive-MIMO systems in realistic UMa propagation environments, the SE was computed using the measured channel matrices $\mathbf{H}$ obtained from the channel sounding campaign described in Section II. The computation follows the Shannon-capacity framework used in the theoretical analysis. Unlike the idealized i.i.d. Rayleigh scenario with full-rank scattering, this empirical evaluation inherently captures the effects of environmental sparsity and spatial correlation of UMa deployments. It is important to note that, in contrast to the theoretical analysis where the receive SNR is fixed through channel normalization, the measured channel matrices are not normalized in this evaluation. Therefore, a constant transmit SNR is assumed, and the effective receive SNR varies across frequency bands due to differences in propagation loss and channel gain.

\begin{figure}[!ht]
\centering
\includegraphics[width=0.9\linewidth]{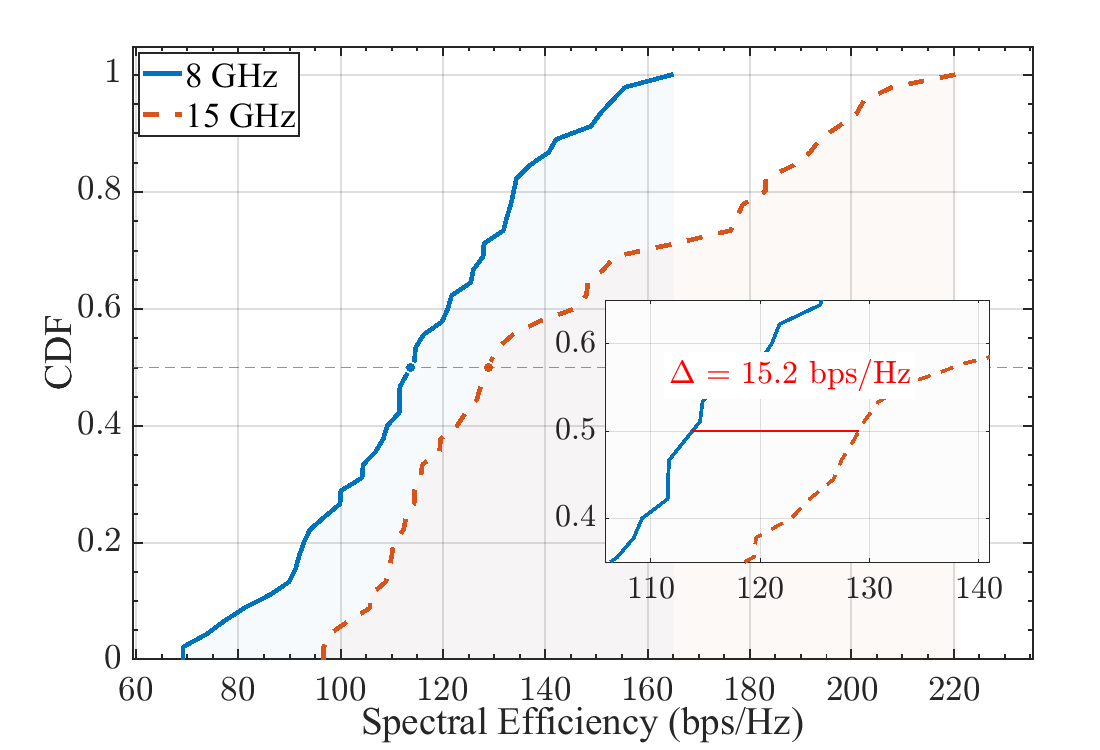}
\caption{Measured spectral efficiency CDFs for 8GHz ($N=32$) and 15GHz ($N=128$) systems under an equal physical aperture constraint. Median values (CDF = 0.5) are highlighted.}
\label{Fig.SE_CDF}
\end{figure}

Fig.~\ref{Fig.SE_CDF} presents the CDF of the measured SE for 8 GHz and 15 GHz systems under an equal physical aperture constraint. At the median (CDF = 0.5), the 8 GHz system with 32 elements achieves an SE of 113.8 bit/s/Hz, whereas the 15 GHz system with 128 elements attains 129.0 bit/s/Hz. This corresponds to a measured SE gain of 15.2 bit/s/Hz, in favor of the 15 GHz system. Table~\ref{tab:SE_compare} further quantifies the comparison, summarizing both the measured SE and the theoretical i.i.d. ceilings, along with the measurement-to-theoretical ratio.

\begin{table}[!ht]
\centering
\caption{Comparison of Empirical and Theoretical Spectral Efficiency}
\label{tab:SE_compare}
\renewcommand{\arraystretch}{1.25}
\footnotesize
\setlength{\tabcolsep}{4pt}
\begin{tabular}{lccc}
\toprule
\textbf{Config.} & \textbf{Measured} & \textbf{Theoretical} & \textbf{Meas./Theo.} \\
($f_c, N_t$)     & (bit/s/Hz)        & (bit/s/Hz)           & \textbf{Ratio} \\
\midrule
8 GHz, $N=32$    & 113.8             & 242.2                & \textbf{47.0\%} \\
8 GHz, $N=128$   & 154.2             & 266.1                & \textbf{58.0\%} \\
15 GHz, $N=128$  & 129.0             & 266.1                & \textbf{48.5\%} \\
\bottomrule
\end{tabular}
\end{table}

A closer inspection of Table~\ref{tab:SE_compare} provides key physical insights. When the element number is matched to 128 for both frequencies, the 8GHz system significantly outperforms the 15~GHz system, achieving 154.2~bps/Hz compared to 129.0~bps/Hz. This clearly demonstrates that the lower-frequency band inherently benefits from richer multipath propagation and lower environmental sparsity, resulting in higher intrinsic SE for the same array elements. In contrast, the measured SE advantage of the 15~GHz system under an equal physical aperture (32 elements at 8~GHz vs. 128 elements at 15~GHz) is primarily a consequence of hardware-enabled spatial multiplexing. The shorter wavelength at 15~GHz permits the deployment of four times as many antenna elements within the same physical area, effectively increasing the system’s spatial degrees of freedom and compensating for the sparser propagation environment. Furthermore, the measurement-to-theoretical ratios provide compelling validation for this physical perspective. Under the equal-aperture constraint, both the 8~GHz ($N=32$) and 15~GHz ($N=128$) systems achieve remarkably similar fractions of their theoretical i.i.d. capacity limits (47.0\% and 48.5\%, respectively). This phenomenon demonstrates that when confined to the same physical footprint, the severity of capacity degradation from the ideal i.i.d. assumption—caused by realistic spatial correlation and limited multipath clusters—is nearly identical for both configurations. 

\subsection{Array Topology Optimization}

Building upon the spectral efficiency evaluation in the previous subsection, we further investigate how the spatial arrangement of antenna elements influences system multiplexing performance. Specifically, three array topologies with identical element counts are synthesized from the measured grid: a $1\times32$ ULA, a $2\times16$ UPA, and a $4\times8$ UPA. This comparison aims to investigate the impact of array geometry and aspect ratio on spectral efficiency while strictly keeping the total hardware RF resources ($N=32$) constant.

\begin{figure}[htbp]
    \centering
    \includegraphics[width=1\columnwidth]{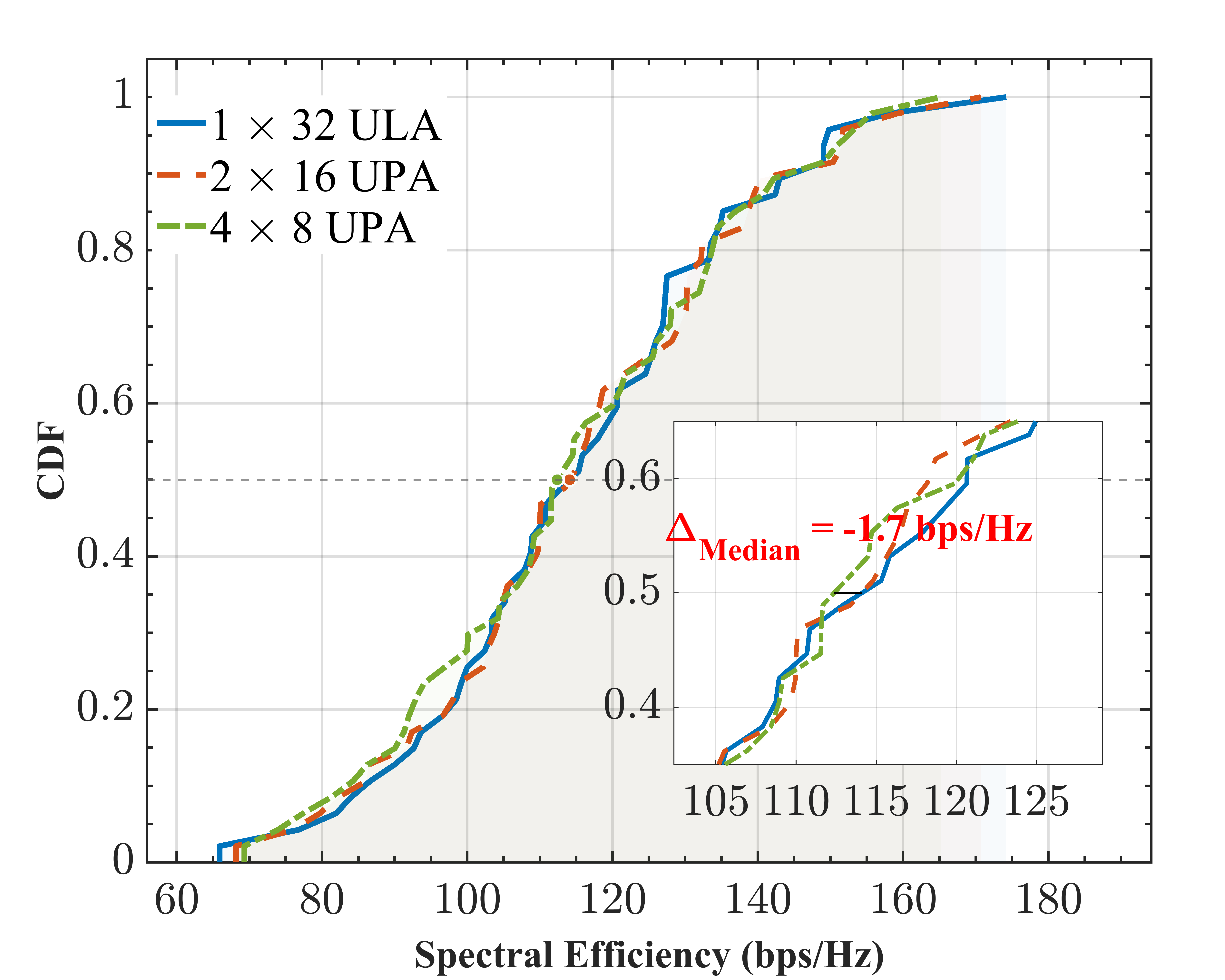}
    \caption{Empirical CDF of real spectral efficiency for different array topologies ($N=32$) at 8~GHz. The results demonstrate that the SE is highly robust to array geometry, with all three configurations achieving nearly identical performance distributions.}
    \label{Fig.SE_topo_CDF}
\end{figure}

Traditionally, it is widely perceived that when the ASA dominates over the ESA, extending the horizontal aperture would enable the system to capture richer angular variations in the dominant plane, thereby enhancing spatial separability and improving spectral efficiency. Under this intuition, array configurations with longer horizontal apertures are often expected to yield superior performance compared to more compact or vertically extended designs. However, the empirical CDFs obtained from measured channels, as shown in Fig.\ref{Fig.SE_topo_CDF}, indicate that this advantage is, in practice, rather limited. While minor performance differences among the three array configurations can indeed be observed, the overall distributions remain closely aligned across the entire probability range. At the median point (CDF = 0.5), the maximum discrepancy between the extreme configurations—the $1\times32$ ULA and the $4\times8$ UPA—is only 1.7bps/Hz, corresponding to a relative difference below 1.5\%. The intermediate $2\times16$ topology follows a nearly identical trend. From a practical deployment perspective, this is a highly encouraging finding. It implies that for Massive-MIMO deployments in typical UMa scenarios, network operators have significant flexibility in designing the physical form factor of the antenna panels. A compact, rectangular $4\times8$ array can be deployed to significantly ease site acquisition, reduce wind-loading, and simplify installation, all without incurring any meaningful penalty in spatial multiplexing capability or overall spectral efficiency compared to a structurally demanding $1\times32$ linear array.

\section{Conclusion}

This paper presented a comprehensive measurement-based study of propagation characteristics and system performance across the 8~GHz and 15~GHz bands in an Urban Macro (UMa) scenario, with a strict focus on comparisons under a fixed physical array aperture constraint. By jointly analyzing large-scale fading, multipath structure, angular dispersion, and system-level metrics, the fundamental impact of frequency scaling on Massive-MIMO channels has been systematically characterized. The results reveal the sparsification in high frequency channel. While the underlying propagation geometry remains largely invariant, higher frequencies experience stronger attenuation, leading to a progressively sparser effective channel. This sparsification manifests across multiple domains, resulting in reduced delay spread, contracted azimuth angular dispersion, and increased sensitivity to local environmental blockages. From a coverage perspective, maintaining a fixed physical aperture enables the 15~GHz system to significantly offset the increased free-space path loss through beamforming. However, this theoretical compensation is incomplete in practical sparse environments. Even under ideal coherent beamforming, a residual coverage deficit of approximately 3.0~dB persists at the cell edge, indicating that aperture-based scaling alone cannot fully overcome the propagation limitations inherent to deep-shadowing conditions at FR3 frequencies. Crucially, the SE evaluation reveals that the capacity advantage of the 15~GHz system under a fixed aperture does not stem from superior intrinsic channel properties. In fact, when the number of array elements is equal ($N=128$), the 8~GHz system significantly outperforms the 15~GHz baseline due to its inherently richer multipath environment. Instead, the observed SE gain at 15~GHz under the equal-aperture constraint is strictly a hardware-enabled dividend. The shorter wavelength allows for integrating four times as many antenna elements within the same physical footprint ($N=128$ vs. $N=32$). Furthermore, contrary to the expectation that maximizing the horizontal aperture is necessary to exploit the dominant azimuth spread in UMa scenarios, empirical data demonstrates that both SE and total captured energy are highly robust to array geometry. In sparse channels, redistributing elements from a highly extended linear array ($1\times32$) into a compact planar structure ($4\times8$) results in negligible performance penalties. 

Overall, these findings indicate that migrating to FR3 introduces a fundamental paradigm shift in Massive-MIMO design. Rather than relying on intrinsic channel richness, high-frequency systems must exploit the spatial hardware dividend enabled by ultra-dense antenna integration. Moreover, the demonstrated robustness of array topology proves that future FR3 base stations can confidently adopt compact, two-dimensional panel designs to ease site acquisition and reduce wind-loading, balancing practical deployment constraints with the stringent coverage and capacity requirements of next-generation networks.

%

\ifCLASSOPTIONcaptionsoff
  \newpage
\fi

\end{document}